\documentclass[english,aip,pop,reprint]{revtex4-1}
\usepackage{lmodern}
\pdfoutput=1

\usepackage[T1]{fontenc}
\usepackage[utf8]{inputenc}
\usepackage{color}
\usepackage{babel}
\usepackage{units}
\usepackage{mathrsfs}
\usepackage{mathtools}
\usepackage{amsmath}
\usepackage{amssymb}
\usepackage{graphicx}
\usepackage[unicode=true,
 bookmarks=true,bookmarksnumbered=false,bookmarksopen=false,
 breaklinks=false,pdfborder={0 0 0},pdfborderstyle={},backref=false,colorlinks=true]
 {hyperref}
\hypersetup{
 pdfauthor={Rudi Gaelzer},
 linkcolor=blue, citecolor=blue}

\makeatletter
\allowdisplaybreaks

\makeatother

\begin{document}
\global\long\def\pFq#1#2{\prescript{\vphantom{#1}}{#1}{F}_{#2}^{\vphantom{#2}}}

\global\long\def\pFRq#1#2{\prescript{\vphantom{#1}}{#1}{\mathfrak{F}}_{#2}^{\vphantom{#2}}}

\title{The oblique firehose instability in a bi-kappa magnetized plasma}

\author{A. R. Meneses}
\email{anemeneses@gmail.com}

\author{R. Gaelzer}
\email{rudi.gaelzer@ufrgs.br}

\author{L. F. Ziebell}
\email{luiz.ziebell@ufrgs.br}

\affiliation{Instituto de Física, UFRGS, 91501-970 Porto Alegre, RS, Brazil}

\begin{abstract}
In this work, we derive a dispersion equation that describes the excitation
of the oblique (or Alfvén) firehose instability in a plasma that contains
both electron and ion species modelled by bi-kappa velocity distribution
functions. The equation is obtained with the assumptions of low-frequency
waves and moderate to large values of the parallel (respective to
the ambient magnetic field) plasma beta parameter, but it is valid
for any direction of propagation and for any value of the particle
gyroradius (or Larmor radius)\@. Considering values for the physical
parameters typical to those found in the solar wind, some solutions
of the dispersion equation, corresponding to the unstable mode, are
presented. In order to implement the dispersion solver, several new
mathematical properties of the special functions occurring in a kappa
plasma are derived and included. The results presented here suggest
that the superthermal characteristic of the distribution functions
leads to reductions to both the maximum growth rate of the instability
and of the spectral range of its occurrence.
\end{abstract}
\maketitle

\section{Introduction}

The plasma environment found in the interplanetary space is, in its
majority, formed by particles of solar origin transported by the solar
wind. The solar wind plasma is composed by electrons, protons, alpha
particles and some other minority ions. Although the measured velocity
distribution functions (VDFs) for each of the major populations have
some specific characteristics, they also feature some common traits,
such as high-energy isotropic or nonisotropic tails and high-energy
beam populations that are aligned with the local interplanetary magnetic
field (IMF).\citep{Marsch06}

One of the most important characteristics shown by the ion VDFs is
a marked and conspicuous anisotropy in the velocity spreads measured
in the direction parallel to the IMF with the spread measured in the
perpendicular direction. These anisotropic velocity spreads are called
in the literature \emph{temperature anisotropies} and are measured
by the second moments of the VDF, respectively evaluated in the parallel
(the \emph{parallel temperature}) and perpendicular (the \emph{perpendicular
temperature}) directions.

The importance of the presence of a temperature anistropy for the
dynamical evolution of the solar wind plasma is that its departure
from a themodynamic equilibrium state means that the plasma contains
free energy sources in the particle distributions that can be tapped
to excite several different plasma instabilities which will be ultimately
responsible for several phenomena observed in the solar wind such
as wave emission, particle energization and turbulence.\citep{KleinHowes15/03}

In spite of the fact that other observed nonequilibrium features such
as particle beams also offer free energy sources, a large part of
the work published in the literature is concerned with the temperature
anisotropy-driven instabilities (TADI)\@. If we restrict ourselves
with the anisotropies caused by the proton VDFs, there are four instabilities
that are usually studied: the electromagnetic ion-cyclotron (or proton
cyclotron) instability (EMIC), roughly excited when $T_{\perp p}>T_{\parallel p}$,
where $T_{\parallel(\perp)p}$ is the parallel (perpendicular) temperature
of the proton VDF, the parallel proton firehose instability (PFH),
excited when $T_{\parallel p}>T_{\perp p}$, the mirror instability
(MI, when $T_{\perp p}>T_{\parallel p}$) and the oblique (or Alfvén)
firehose instability (OFH, when $T_{\parallel p}>T_{\perp p}$)\@.\citep{KleinHowes15/03}
These instabilities are grouped in such a way that the first two (EMIC
and PFH) are usually studied in the direction parallel to the IMF,
whereas the second couple (MI and OFH) occur in the oblique direction.
Rather than giving a long list of publications on the subject, the
Reader is referred to the recent review provided by Ref. \onlinecite{Yoon17/12},
which also provides lengthier discussions both on the theoretical
derivation of the instabilities from the kinetic theory of plasmas
and about their importance for the physical processes that take place
in the solar wind.

The vast majority of the theoretical work on the TADI has been done
assuming that the VDFs of the solar wind species can be adequately
fitted by a combination of Maxwellian and bi-Maxwellian distributions.\citep{KleinHowes15/03,Yoon17/12,Klein+18/05}
However, a substantial fraction of observed VDFs display high-energy
(\emph{superthermal}) tails that are better fitted by some power-law
dependence such as $f\left(\boldsymbol{v}\right)\propto v^{-2\kappa}$,
where $f\left(\boldsymbol{v}\right)$ is the velocity distribution
function and $0<\kappa<\infty$ is a fitting parameter. 

The most frequently employed model of distribution with such power-law
behavior is the kappa VDF,\citep{Livadiotis15/03} which describes
the velocity distribution of a plasma species that is in a quasi-stationary
state away from thermal equilibrium, where the particle interactions
are long-ranged and where there are strong correlations among the
degrees of freedom. With a $\kappa$VDF, the kappa parameter is a
measure of the departure of the (quasi-)stationary state from thermal
equilibrium: the smaller the value of $\kappa$ the farther from equilibrium,
which is asymptotically reached when $\kappa\to\infty$\@. In practice,
when $\kappa\gtrsim20$ the $\kappa$VDF is already quasi-Maxwellian.
A collection of works regarding the origin, observations, properties
and the statistical mechanics of kappa distributions was recently
organized by G. Livadiotis.\citep{Livadiotis17}

It is already well established that the high energy populations of
the electron VDF in the solar wind (the halo and the Strahl) are better
modelled with kappa or bi-kappa distributions.\citep{Maksimovic+05/09,Stverak+09/05,Pierrard+16/08}
For both populations, it has been measured $\kappa_{e}<10$ for a
wide range of heliocentric distances, ranging from 0.3 AU (Astronomical
Units) to nearly 4 AU\@. In fact, the value of $\kappa_{e}$ reduces
with distance, showing that the electrons in the solar wind are in
a constant process of departure from thermal equilibrium as the solar
wind propagates through the interplanetary space. 

The observed distributions of the major ion species in the solar wind
and some physical properties associated with the shape of the VDFs
have also been analysed employing kappa distributions. Detailed discussions
and a longer list of publications can be found in Refs. \onlinecite{PierrardPieters14/12,Livadiotis+18/02}\@.
It has been verified that the ion VDFs are also well modelled with
typical values of $\kappa_{i}\lesssim5$\@.\footnote{See also table 1.1 of Livadiotis\cite{Livadiotis17c} for measured
values of $\kappa$ in other solar and astrophysical environments.} 

Therefore, one must conclude that a full picture of the plasma instabilities
operating in the solar wind can only be obtained if, in their theoretical
description, the superthermal nature of the observed distributions
are taken into account.

So far, of the four instabilities listed above, excited by temperature
anisotropies in the ion VDFs, only those that occur in parallel-propagating
modes (MTSI and PFH) have been systematicaly studied when $\kappa$
distributions are involved.\citep{Lazar+11/10,Lazar12/11,LazarPoedts14/01,dosSantos+14/11,dosSantos+15/12,dosSantos+16/01,dosSantos+17/01,ZiebellGaelzer17/09,Lazar+18/02}
The Reader is referred to a recent review by Viñas \emph{et al.}\citep{Vinas+17}
for a detailed theoretical account and further references.

By comparison, studies regarding instabilities that are excited by
$\kappa$VDFs and that occur at oblique angles relative to the IMF
are scant. One line of research has been focused on the development
of computer codes that numerically evaluate the dielectric tensor
of kappa distributions and solve the dispersion equation, thereby
obtaining the dispersion relations of the several wavemodes and their
associated damping/growth rates with an \emph{ab initio} numerical
procedure.\citep{Sugiyama+15/10,Astfalk+15/09,AstfalkJenko16/12} 

Another line has emphasized the derivation of analytical and closed-form
expressions for the dielectric tensor of a (bi-)kappa VDF\@. A first
contribution  considered the effect of the $\kappa$VDF on the propagation
and damping of highly-oblique dispersive Alfvén waves in the Earth's
magnetosphere\@.\citep{GaelzerZiebell14/12} Further contributions
provided the closed-form expressions for the dielectric tensor of
an isotropic kappa plasma in Ref. \onlinecite{GaelzerZiebell16/02}
and a bi-kappa plasma in Ref. \onlinecite{Gaelzer+16/06}\@. These
publications will hereafter be addressed respectively as Papers I
and II\@. Whenever possible, the derivation of analytical and closed-form
expressions for the dielectric tensor is desirable, because they provide
important information about the mathematical properties of the dispersion
relations and allow the derivation of the dispersion equations for
specific normal modes of propagation.

In this work, we will employ the formulation provided in Papers I
and II in order to derive a dispersion equation suitable for the study
of the oblique firehose instability excited by a bi-kappa plasma.
The OFH, first discovered by Yoon \emph{et al.}\citep{YoonWuAssis93/07}
and later found again by Hellinger and Matsumoto,\citep{HellingerMatsumoto00/05}
is an instability occurring in low-frequencies and at oblique angles,
commonly associated with dispersive Alfvén waves.\citep{LysakLotko96/03,GaelzerZiebell14/12}
However, as we shall see below, the OFH is in fact an absolute instability
associated with a nonpropagating mode. Here, we will derive a dispersion
equation valid for plasmas with moderate values of the ion beta parameter
and show some numerical results. In order to implement the dispersion
relation solver, several new mathematical properties of the special
functions ocurring in a kappa plasma were obtained, which were not
presented in Papers I and II\@. Consequently, in this work we will
present some typical solutions of the dispersion equation and reserve
a more comprehensive and detailed analysis of the instability for
a future publication.

The plan of the paper is as follows. In section \ref{sec:MGZ18:DE}
the specific dispersion equation is obtained. In section \ref{sec:MGZ18:Kappa_plasma_functions},
the new mathematical properties for the special functions are derived.
Then, in section \ref{sec:MGZ18:Numerical_sols} we present some solutions
of the dispersion equation, which show the occurrence of the OFH\@.
Finally, in section \ref{sec:MGZ18:Conclusions} we present our conclusions.

\section{Dispersion equation for a high-beta bi-kappa plasma\label{sec:MGZ18:DE}}

The general form of the dielectric tensor for a magnetized bi-kappa
plasma can be found in Paper II, eqs. 2 and 3\@. For the present
application, we will employ the kappa velocity distribution function
($\kappa$VDF) introduced first by Summers and Thorne,\citep{SummersThorne91/08}
which can be obtained from the general form adopted in Papers I and
II by setting the parameters $\alpha=1$ and $w_{\parallel(\perp)s}^{2}=\left(1-3/2\kappa_{s}\right)v_{T\parallel(\perp)s}^{2}$,
where $s=e,i,\dots$ denotes the plasma species/population and $v_{T\parallel(\perp)s}^{2}=2T_{\parallel(\perp)s}/m_{s}$
is the thermal velocity squared of the same species. With this particular
choice, the explicit expression of the $\kappa$VDF is obtained from
(I.1) (\emph{i.e.}, equation 1 of Paper I) or from (II.1) and is given
by
\begin{align}
f_{s}\left(\boldsymbol{v}\right) & =A_{s}^{(\kappa_{s})}\left(1+\frac{v_{\parallel}^{2}}{\kappa_{s}w_{\parallel s}^{2}}+\frac{v_{\perp}^{2}}{\kappa_{s}w_{\perp s}^{2}}\right)^{-\left(\kappa_{s}+1\right)}\label{eq:MGZ18:kVDF}\\
A_{s}^{(\kappa_{s})} & =\frac{1}{\pi^{3/2}w_{\parallel s}w_{\perp s}^{2}}\frac{\Gamma\left(\kappa_{s}+1\right)}{\kappa_{s}^{3/2}\Gamma\left(\kappa_{s}-\nicefrac{1}{2}\right)},\:\left(\kappa_{s}>\frac{1}{2}\right).\nonumber 
\end{align}

In accordance with the particular choices of the parameters $\alpha$
and $w_{\parallel(\perp)s}$ adopted here, the same settings must
the imposed on the general expressions for the dielectric tensor and
the kappa plasma functions discussed in Papers I and II.

The particular form of the $\kappa$VDF in (\ref{eq:MGZ18:kVDF})
is the model more frequently employed in the literature for reasons
that have been discussed at length in Ref. \onlinecite{GaelzerZiebell14/12}
and in Papers I and II.

Among the several new expressions introduced by Papers I and II regarding
the physical properties of the propagation of electromagnetic/electrostatic
waves and wave-particle interactions in a kappa plasma, for any combination
of particle species and wave modes, frequency and polarization, a
great amount of space was dedicated to the discussion of the mathematical
properties and numerical evaluation of several special functions,
namely, the \emph{kappa }(or \emph{superthermal}) \emph{plasma dispersion
function }$Z_{\kappa}^{\left(\alpha,\beta\right)}\left(\xi\right)$,
the \emph{kappa plasma gyroradius function }$\mathcal{H}_{n,\kappa}^{\left(\alpha,\beta\right)}\left(z\right)$
and the \emph{two-variables kappa plasma functions }$\mathcal{Z}_{n,\kappa}^{\left(\alpha,\beta\right)}\left(\mu,\xi\right)$
and $\mathcal{Y}_{n,\kappa}^{\left(\alpha,\beta\right)}\left(\mu,\xi\right)$\@.
The reader is referred to Papers I and II for the definitions of these
functions and for the mentioned properties. Some new mathematical
formulas and properties for the same functions, developed for the
present application, are presented in section \ref{sec:MGZ18:Kappa_plasma_functions}
below.

In order to reduce the amount of algebra, we will first define the
quantity 
\[
1_{\kappa}^{\left(\beta\right)}=\frac{\Gamma\left(\kappa+\beta-1\right)}{\kappa^{\beta-1/2}\Gamma\left(\kappa-\nicefrac{1}{2}\right)},
\]
which is such that $1_{\kappa}^{\left(\beta\right)}\xrightarrow{\kappa\to\infty}1$,
and then define the following form for the kappa plasma gyroradius
function, 
\[
\mathrm{H}_{n,\kappa}^{\left(\beta\right)}\left(\mu\right)=1_{\kappa}^{\left(\beta+1\right)}\mathcal{H}_{n,\kappa}^{\left(1,\beta\right)}\left(\mu\right).
\]
We will also employ hereafter the shorthand notation $\mathcal{Z}_{n,\kappa}^{\left(\beta\right)}\left(\mu,\xi\right)$
and $\mathcal{Y}_{n,\kappa}^{\left(\beta\right)}\left(\mu,\xi\right)$,
which implicitly assumes that we are taking the same functions with
$\alpha=1$.

Given a plasma species $s$ composed by particles with mass $m_{s}$,
electric charge $q_{s}$ and number density $n_{s}$, we start by
taking the general form of the dielectric tensor in (II.2,3) and make
the change $n\to-n$ for all the terms involving negative values of
the harmonic index $n$, thereby obtaining\begin{subequations}\label{eq:MGZ18:DT-BK-HB}
\begin{align}
\varepsilon_{xx} & =1+\sum_{s}\frac{\omega_{ps}^{2}}{\omega^{2}}\sum_{n=1}^{\infty}\frac{n^{2}}{\mu_{s}}\left(A_{ns,\kappa}+A_{-ns,\kappa}\right)\\
\varepsilon_{xy} & =i\sum_{s}\frac{\omega_{ps}^{2}}{\omega^{2}}\sum_{n=1}^{\infty}n\partial_{\mu_{s}}\left(A_{ns,\kappa}-A_{-ns,\kappa}\right)\\
\varepsilon_{xz} & =-\sum_{s}\frac{\omega_{ps}^{2}}{\omega^{2}}\frac{k_{\perp}w_{\parallel s}}{2\Omega_{s}}\sum_{n=1}^{\infty}\frac{n}{\mu_{s}}\left(B_{ns,\kappa}-B_{-ns,\kappa}\right)\\
\varepsilon_{yy} & =1+\sum_{s}\frac{\omega_{ps}^{2}}{\omega^{2}}\left[C_{0s,\kappa}+\sum_{n=1}^{\infty}\left(C_{ns,\kappa}+C_{-ns,\kappa}\right)\right]\\
\varepsilon_{yz} & =i\sum_{s}\frac{\omega_{ps}^{2}}{\omega^{2}}\frac{k_{\perp}w_{\parallel s}}{2\Omega_{s}}\left[\partial_{\mu_{s}}B_{0s,\kappa}\right.\nonumber \\
 & \left.+\sum_{n=1}^{\infty}\partial_{\mu_{s}}\left(B_{ns,\kappa}+B_{-ns,\kappa}\right)\right]\\
\varepsilon_{zz} & =1-\sum_{s}\frac{\omega_{ps}^{2}}{\omega^{2}}\frac{v_{T\parallel s}^{2}}{v_{T\perp s}^{2}}\left[\xi_{0s}B_{0s,\kappa}\right.\nonumber \\
 & \left.+\sum_{n=1}^{\infty}\left(\xi_{ns}B_{ns,\kappa}+\xi_{-ns}B_{-ns,\kappa}\right)\right],
\end{align}
where $\varepsilon_{ij}$ $\left(i,j=x,y,z\right)$ are the components
of the dielectric tensor, which obeys the usual symmetry relations
$\varepsilon_{xy}=-\varepsilon_{yx}$, $\varepsilon_{xz}=\varepsilon_{zx}$
and $\varepsilon_{yz}=-\varepsilon_{zy}$\@. Also, $\partial_{\mu}=\partial/\partial\mu$,
\begin{align*}
\mu_{s} & =\frac{k_{\perp}^{2}w_{\perp s}^{2}}{2\Omega_{s}^{2}}=\left(1-\frac{3}{2\kappa_{s}}\right)\frac{k_{\perp}^{2}v_{T\perp s}^{2}}{2\Omega_{s}^{2}},\\
\xi_{ns} & =\frac{\omega-n\Omega_{s}}{k_{\parallel}w_{\parallel s}}=\left(1-\frac{3}{2\kappa_{s}}\right)^{-\frac{1}{2}}\frac{\omega-n\Omega_{s}}{k_{\parallel}v_{T\parallel s}},
\end{align*}
and $\omega_{ps}^{2}=4\pi n_{s}q_{s}^{2}/m_{s}$ is the squared plasma
and $\Omega_{s}=q_{s}B_{0}/m_{s}c$ the cyclotron frequencies of species
$s$ for a plasma embedded in an uniform ambient magnetic induction
vector $\boldsymbol{B}_{0}=B_{0}\hat{\boldsymbol{z}}$\@. Additionaly,
$c$ is the light speed in vacuum and $\omega$ and $\boldsymbol{k}=k_{\perp}\hat{\boldsymbol{x}}+k_{\parallel}\hat{\boldsymbol{z}}$
are the usual (angular) frequency and wave vector of oscillations
propagating in the plasma.

\end{subequations}

In (\ref{eq:MGZ18:DT-BK-HB}a-f), the quantities $A$, $B$ and $C$
are defined as\begin{subequations}\label{eq:MGZ18:Coefs-ABC}
\begin{align}
A_{ns,\kappa} & =\xi_{0s}\mathcal{Z}_{n,\kappa_{s}}^{\left(2\right)}\left(\mu_{s},\xi_{ns}\right)+\frac{1}{2}A_{s}\partial_{\xi_{ns}}\mathcal{Z}_{n,\kappa_{s}}^{\left(1\right)}\left(\mu_{s},\xi_{ns}\right)\\
B_{ns,\kappa} & =\left(\xi_{0s}-A_{s}\xi_{ns}\right)\partial_{\xi_{ns}}\mathcal{Z}_{n,\kappa_{s}}^{\left(1\right)}\left(\mu_{s},\xi_{ns}\right)\\
C_{ns,\kappa} & =\xi_{0s}\mathcal{W}_{n,\kappa_{s}}^{\left(2\right)}\left(\mu_{s},\xi_{ns}\right)+\frac{1}{2}A_{s}\partial_{\xi_{ns}}\mathcal{W}_{n,\kappa_{s}}^{\left(1\right)}\left(\mu_{s},\xi_{ns}\right),
\end{align}
where 
\[
\mathcal{W}_{n,\kappa_{s}}^{\left(\beta\right)}\left(\mu_{s},\xi_{ns}\right)=\frac{n^{2}}{\mu_{s}}\mathcal{Z}_{n,\kappa_{s}}^{\left(\beta\right)}\left(\mu_{s},\xi_{ns}\right)-2\mu_{s}\mathcal{Y}_{n,\kappa_{s}}^{\left(\beta\right)}\left(\mu_{s},\xi_{ns}\right),
\]
 $A_{s}=1-T_{\perp s}/T_{\parallel s}$ is the temperature anisotropy
parameter for species $s$, and $\partial_{\xi}=\partial/\partial\xi$.\end{subequations}

In the present application, we are interested in low-frequency $(\omega\ll\left|\Omega_{s}\right|)$
and long-wavelength $\bigl(k_{\parallel}^{2}v_{T\parallel s}^{2}\ll\Omega_{s}^{2}\bigr)$
waves propagating in oblique directions relative to $\boldsymbol{B}_{0}$\@.
In such a situation, $\omega_{ps}^{2}/\omega^{2}\gg\omega_{ps}^{2}/\Omega_{s}^{2}$,
where 
\[
\frac{\omega_{ps}^{2}}{\Omega_{s}^{2}}=\frac{n_{s}m_{s}}{n_{i}m_{i}}\frac{c^{2}}{v_{A}^{2}},
\]
with $s=i$ referring to the ion species and with the squared Alfvén
speed $v_{A}^{2}=B_{0}^{2}/4\pi n_{i}m_{i}$\@. Hence, for an electron-ion
plasma, if at least we have $\left(m_{e}/m_{i}\right)\left(c^{2}/v_{A}^{2}\right)\simeq1$,
then $\omega_{ps}^{2}/\omega^{2}\gg1$ and we can neglect, as a first
approximation, the displacement current terms (the unity) in the diagonal
components of the dielectric tensor (\ref{eq:MGZ18:DT-BK-HB}).

Now we notice that given the (squared) particle gyroradius (or Larmor
radius) $\rho_{s}^{2}=v_{T\perp s}^{2}/2\Omega_{s}^{2}=T_{\perp s}/m_{s}\Omega_{s}^{2}$,
the quantity $\mu_{s}$ can be written as $\mu_{s}=\left(1-3/2\kappa_{s}\right)k_{\perp}^{2}\rho_{s}^{2}$\@.
Among the several low-frequency wave modes observed in the solar wind
and Earth's magnetosphere, of particular importance are the dispersive
Alfvén waves (DAW)\@.\citep{LysakLotko96/03,GaelzerZiebell14/12}
In regions where the plasma thermal effects on the wave dispersion
can not be ignored, the DAW are known as the kinetic Alfvén waves
(KAW)\@. Measuring the thermal effect with the parallel/perpendicular
plasma beta parameter $\beta_{\parallel(\perp)s}=8\pi n_{s}T_{\parallel(\perp)s}/B_{0}^{2}$,
the dispersion relation of KAW propagating in an isotropic Maxwellian
plasma with moderate values of the electron beta $\left(m_{e}/m_{i}\ll\beta_{e}\lesssim1\right)$
is\citep{LysakLotko96/03}
\[
\frac{\omega^{2}}{k_{\parallel}^{2}v_{A}^{2}}\approx1+k_{\perp}^{2}\left(\rho_{a}^{2}+\frac{3}{4}\rho_{i}^{2}\right),
\]
where $\rho_{a}^{2}=T_{e}/m_{i}\Omega_{i}^{2}$ is the ion-acoustic
gyroradius. Hence, for the ions, an estimate on the magnitude of the
parameter $\mu_{i}$ for KAW is given by 
\[
\mu_{i}=\frac{1}{2}\left(1-\frac{3}{2\kappa_{i}}\right)\frac{k_{\perp}^{2}c^{2}}{\omega_{pi}^{2}}\beta_{\perp i}\approx\frac{1}{2}\left(1-\frac{3}{2\kappa_{i}}\right)\frac{\omega^{2}}{\Omega_{i}^{2}}\frac{k_{\perp}^{2}}{k_{\parallel}^{2}}\beta_{\perp i}.
\]

Therefore, even for low-frequency Alfvén waves, the parameter $\mu_{i}$
can be small but finite when the ion beta is large and/or the KAW
is propagating at large angles relative to $\boldsymbol{B}_{0}$.

Moreover, since we are assuming $\omega\ll\left|\Omega_{s}\right|$
and $k_{\parallel}^{2}v_{T\parallel s}^{2}\ll\Omega_{s}^{2}$, the
typical $\left|\xi_{ns}\right|$ for $n\geqslant1$ is supposed to
be larger than the unity by several orders of magnitude. In this situation,
we can employ in the coefficients $A_{ns,\kappa}$, $B_{ns,\kappa}$
and $C_{ns,\kappa}$ the asymptotic expansions given in (II.28a,b)
for the functions $\mathcal{Z}_{n,\kappa}^{\left(\beta\right)}\left(\mu,\xi\right)$
and $\mathcal{Y}_{n,\kappa}^{\left(\beta\right)}\left(\mu,\xi\right)$,
with their derivatives given by (II.22) and (\ref{eq:MGZ18:dY_cal-dxi}),
keeping only the leading terms in the expansions. Then, defining the
small parameters $x=\omega/n\Omega_{s}$ and $\epsilon=k_{\parallel}w_{\parallel s}/n\Omega_{s}$
for $n\geqslant1$, expanding the combinations of the coefficients
in (\ref{eq:MGZ18:DT-BK-HB}) and (\ref{eq:MGZ18:Coefs-ABC}) in powers
of $x$ and $\epsilon$ and keeping only the lowest-order contributions,
one obtains, after some algebra,
\begin{align*}
A_{ns,\kappa}+A_{-ns,\kappa} & \approx\frac{2\omega^{2}}{n^{2}\Omega_{s}^{2}}\left[\vphantom{\frac{k_{\parallel}^{2}}{\omega^{2}}}\mathrm{H}_{n,\kappa_{s}}^{\left(\nicefrac{3}{2}\right)}\left(\mu_{s}\right)\right.\\
 & \left.+\frac{1}{2}A_{s}\mathrm{H}_{n,\kappa_{s}}^{\left(\nicefrac{1}{2}\right)}\left(\mu_{s}\right)\frac{k_{\parallel}^{2}w_{\parallel s}^{2}}{\omega^{2}}\right]\\
A_{ns,\kappa}-A_{-ns,\kappa} & \approx\frac{2\omega}{n\Omega_{s}}\left[\vphantom{\frac{k_{\parallel}^{2}}{\omega^{2}}}\mathrm{H}_{n,\kappa_{s}}^{\left(\nicefrac{3}{2}\right)}\left(\mu_{s}\right)\right.\\
 & \left.+\frac{1}{2}\left(1+2A_{s}\right)\mathrm{H}_{n,\kappa_{s}}^{\left(\nicefrac{1}{2}\right)}\left(\mu_{s}\right)\frac{k_{\parallel}^{2}w_{\parallel s}^{2}}{n^{2}\Omega_{s}^{2}}\right]\\
B_{ns,\kappa}+B_{-ns,\kappa} & \approx2\left(A_{s}+1\right)\mathrm{H}_{n,\kappa_{s}}^{\left(\nicefrac{1}{2}\right)}\left(\mu_{s}\right)\frac{k_{\parallel}w_{\parallel s}}{n\Omega_{s}}\frac{\omega}{n\Omega_{s}}\\
B_{ns,\kappa}-B_{-ns,\kappa} & \approx2\mathrm{H}_{n,\kappa_{s}}^{\left(\nicefrac{1}{2}\right)}\left(\mu_{s}\right)A_{s}\frac{k_{\parallel}w_{\parallel s}}{n\Omega_{s}}\\
C_{ns,\kappa}+C_{-ns,\kappa} & \approx2\mathrm{G}_{n,\kappa_{s}}^{\left(\nicefrac{3}{2}\right)}\left(\mu_{s}\right)\frac{\omega^{2}}{n^{2}\Omega_{s}^{2}}\\
 & +A_{s}\mathrm{G}_{n,\kappa_{s}}^{\left(\nicefrac{1}{2}\right)}\left(\mu_{s}\right)\frac{k_{\parallel}^{2}w_{\parallel s}^{2}}{n^{2}\Omega_{s}^{2}}
\end{align*}
\[
\xi_{ns}B_{ns,\kappa}+\xi_{-ns}B_{-ns,\kappa}\approx-2\mathrm{H}_{n,\kappa}^{\left(\nicefrac{1}{2}\right)}\left(\mu_{s}\right)A_{s},
\]
where 
\[
\mathrm{G}_{n,\kappa}^{\left(\beta\right)}\left(\mu\right)=\frac{n^{2}}{\mu}\mathrm{H}_{n,\kappa}^{\left(\beta\right)}\left(\mu\right)-2\mu\mathrm{H}_{n,\kappa}^{\left(\beta-1\right)\prime}\left(\mu\right).
\]

Notice that the above approximations are only used for the terms with
harmonic $n\neq0$\@. In the expressions for $A_{0s,\kappa}$, $B_{0s,\kappa}$
and $C_{0s,\kappa}$, we have kept the full thermal effects for the
kappa plasma functions. 

Inserting these approximations back into (\ref{eq:MGZ18:DT-BK-HB}),
we observe that each component of the dielectric tensor contains a
sum over $n\geqslant1$ of the function $\mathrm{H}_{n,\kappa}^{\left(\beta\right)}\left(\mu\right)$
or its derivative. Making use of the identity (II.10), we can establish
the sum rule
\[
\sum_{n\to-\infty}^{\infty}\mathrm{H}_{n,\kappa}^{\left(\beta\right)}\left(\mu\right)=1_{\kappa}^{\left(\beta\right)},
\]
whereby we can define the auxiliary functions \begin{subequations}\label{eq:MGZ18:H1-H2}
\begin{align}
H_{1,\kappa}^{\left(\beta\right)}\left(\mu\right) & =\frac{1_{\kappa}^{\left(\beta\right)}-\mathrm{H}_{0,\kappa}^{\left(\beta\right)}\left(\mu\right)}{\mu},\\
H_{2,\kappa}^{\left(\beta\right)}\left(\mu\right) & =2\sum_{n=1}^{\infty}\frac{\mathrm{H}_{n,\kappa}^{\left(\beta\right)\prime}\left(\mu\right)}{n^{2}},
\end{align}
which were obtained by also making use of the property $\mathrm{H}_{-n,\kappa}^{\left(\beta\right)}\left(\mu\right)=\mathrm{H}_{n,\kappa}^{\left(\beta\right)}\left(\mu\right)$\@.
Additionaly, by virtue of (I.21) and the identity found at the top
of page 4 of Paper II, we have $H_{1,\kappa}^{\left(\beta\right)}\left(0\right)=H_{2,\kappa}^{\left(\beta\right)}\left(0\right)=1_{\kappa}^{(\beta-1)}$.
\end{subequations}

Apropos, the Maxwellian limits of functions $H_{1,\kappa}^{\left(\beta\right)}$
and $H_{2,\kappa}^{\left(\beta\right)}$ are
\begin{align*}
H_{1}\left(\mu\right) & =\frac{1-\mathscr{H}_{0}\left(\mu\right)}{\mu}=1-\frac{3}{4}\mu+\frac{5}{12}\mu^{2}+\cdots\\
H_{2}\left(\mu\right) & =2\sum_{n=1}^{\infty}\frac{\mathscr{H}_{n}^{\prime}\left(\mu\right)}{n^{2}}=1-\frac{15}{8}\mu+\frac{245}{144}\mu^{2}+\cdots,
\end{align*}
where $\mathscr{H}_{n}\left(\mu\right)=e^{-\mu}I_{n}\left(\mu\right)$\@.
The small gyroradius expansion of $H_{1}\left(\mu\right)$ presented
above is well known and easily obtained from the properties of the
modified Bessel function. However, the function $H_{2}\left(\mu\right)$
has always been presented in closed form and, to the best of the authors'
knowledge, no series expansion was known in the literature (see, \emph{e.g.},
Refs. \onlinecite{YoonWuAssis93/07,LysakLotko96/03})\@. The derivation
of the series expansion for $H_{2}\left(\mu\right)$ will be given
in a future publication.

Finaly, considering a 2-species electron-ion plasma, we define the
following parameters, which are inspired by and generalize the corresponding
parameters given in Ref. \onlinecite{YoonWuAssis93/07}, \begin{subequations}\label{eq:MGZ18:DT-BK-pars}
\begin{align}
\epsilon_{\kappa}^{\left(\beta\right)} & =H_{1,\kappa_{i}}^{\left(\beta\right)}\left(\mu_{i}\right)\\
\epsilon_{\kappa}^{\left(\beta\right)\prime} & =\epsilon_{\kappa}^{\left(\beta\right)}-2\mu_{i}H_{2,\kappa_{i}}^{\left(\beta-1\right)}\left(\mu_{i}\right)\\
\hat{\epsilon}_{\kappa} & =-\left[\frac{Z_{\kappa_{e}}^{\left(0\right)\prime}\left(\xi_{0e}\right)}{\beta_{\parallel e,\kappa}}+\frac{1}{\beta_{\parallel i,\kappa}}\partial_{\xi_{0i}}\mathcal{Z}_{0,\kappa_{i}}^{\left(1\right)}\left(\mu_{i},\xi_{0i}\right)\right]\frac{\beta_{\perp i,\kappa}}{2\mu_{i}}\\
\gamma_{\kappa} & =\frac{1}{2}\left(\beta_{\parallel e}A_{e}+\beta_{\parallel i,\kappa}A_{i}\epsilon_{\kappa}^{\left(\nicefrac{1}{2}\right)}-2\right)\frac{k_{\parallel}^{2}v_{A}^{2}}{\Omega_{i}^{2}}\\
\gamma_{\kappa}^{\prime} & =\frac{1}{2}\left(\beta_{\parallel e}A_{e}+\beta_{\parallel i,\kappa}A_{i}\epsilon_{\kappa}^{\left(\nicefrac{1}{2}\right)\prime}-2\right)\frac{k_{\parallel}^{2}v_{A}^{2}}{\Omega_{i}^{2}}\nonumber \\
 & -\left[1+\beta_{\perp e,\kappa}\left(1_{\kappa_{e}}^{\left(-\nicefrac{1}{2}\right)}+\frac{1}{2}\frac{\beta_{\perp e,\kappa}}{\beta_{\parallel e,\kappa}}Z_{\kappa_{e}}^{\left(-2\right)\prime}\left(\xi_{0e}\right)\right)\right.\nonumber \\
 & -\beta_{\perp i,\kappa}\left(\vphantom{\frac{\beta_{\perp i,\kappa}}{\beta_{\parallel i,\kappa}}}\mathrm{H}_{0,\kappa_{i}}^{\left(\nicefrac{1}{2}\right)\prime}\left(\mu_{i}\right)\right.\nonumber \\
 & \left.\left.+\frac{1}{2}\frac{\beta_{\perp i,\kappa}}{\beta_{\parallel i,\kappa}}\partial_{\xi_{0i}}\mathcal{Y}_{0,\kappa_{i}}^{\left(1\right)}\left(\mu_{i},\xi_{0i}\right)\right)\right]\frac{2\mu_{i}}{\beta_{\perp i,\kappa}}\\
\eta_{\kappa} & =1_{\kappa_{e}}^{\left(\nicefrac{1}{2}\right)}+\mathrm{H}_{0,\kappa_{i}}^{\left(\nicefrac{3}{2}\right)\prime}\left(\mu_{i}\right)\nonumber \\
 & -\frac{1}{2}\left(1+2A_{i}\right)\beta_{\parallel i,\kappa}H_{2,\kappa_{i}}^{\left(\nicefrac{1}{2}\right)}\left(\mu_{i}\right)\frac{k_{\parallel}^{2}v_{A}^{2}}{\Omega_{i}^{2}}\\
\eta_{\kappa}^{\prime} & =-\frac{\beta_{\perp e,\kappa}}{2\beta_{\parallel e,\kappa}}Z_{\kappa_{e}}^{\left(-1\right)\prime}\left(\xi_{0e}\right)-\frac{\beta_{\perp i,\kappa}}{2\beta_{\parallel i,\kappa}}\partial_{\mu_{i}}\partial_{\xi_{0i}}\mathcal{Z}_{0,\kappa_{i}}^{\left(1\right)}\left(\mu_{i},\xi_{0i}\right)\nonumber \\
 & -\frac{1}{2}\beta_{\parallel i,\kappa}\left(1+A_{i}\right)H_{2,\kappa_{i}}^{\left(\nicefrac{1}{2}\right)}\left(\mu_{i}\right)\frac{k_{\parallel}^{2}v_{A}^{2}}{\Omega_{i}^{2}},
\end{align}
where we have also defined the kappa-modified beta parameters as
\[
\beta_{\parallel(\perp)s,\kappa}=\left(1-\frac{3}{2\kappa_{s}}\right)\beta_{\parallel(\perp)s}=\frac{\beta_{\parallel(\perp)s}}{1_{\kappa_{s}}^{(-\nicefrac{1}{2})}}.
\]
We must also point out that since $\mu_{e}\ll\mu_{i}$ by a factor
of order $m_{e}/m_{i}$, we have kept in parameters (\ref{eq:MGZ18:DT-BK-pars}a-g)
the lowest-order gyroradius contribution from the electrons, \emph{i.e.},
 $H_{1,\kappa_{e}}^{\left(\beta\right)}\left(\mu_{e}\right)\approx H_{2,\kappa_{e}}^{\left(\beta\right)}\left(\mu_{e}\right)\approx1_{\kappa_{e}}^{(\beta-1)}$.\end{subequations}

The parameters (\ref{eq:MGZ18:DT-BK-pars}a-g) are identified within
the dispersion equation $\det\mathsf{\Lambda}=0$, where the elements
of the matrix $\mathsf{\Lambda}$ are 
\[
\Lambda_{ij}=\frac{c^{2}}{\omega^{2}}\left(k_{i}k_{j}-k^{2}\delta_{ij}\right)+\boldsymbol{\varepsilon}_{ij},\;\left(i,j=x,y,z\right).
\]
After some algebraic manipulation, during the course of which some
other terms that are of order $m_{e}/m_{i}$ are neglected, one notices
that one root $\omega^{2}\approx0$ can be factored out and the remaining
equation can be written in transcendental form as 
\begin{multline}
\left(\frac{\omega}{\Omega_{i}}\right)^{4}+\left[\frac{1}{\epsilon_{\kappa}^{\left(\nicefrac{3}{2}\right)\prime}}\left(\gamma_{\kappa}^{\prime}-\frac{\eta_{\kappa}^{2}}{\epsilon_{\kappa}^{\left(\nicefrac{3}{2}\right)}}-\frac{\eta_{\kappa}^{\prime2}}{\hat{\epsilon}_{\kappa}}\right)\right.\\
\left.+\gamma_{\kappa}\left(\frac{1}{\epsilon_{\kappa}^{\left(\nicefrac{3}{2}\right)}}+\frac{1}{\hat{\epsilon}_{\kappa}}\right)\right]\left(\frac{\omega}{\Omega_{i}}\right)^{2}+\frac{\gamma_{\kappa}}{\epsilon_{\kappa}^{\left(\nicefrac{3}{2}\right)\prime}}\\
\times\left[\gamma_{\kappa}^{\prime}\left(\frac{1}{\epsilon_{\kappa}^{\left(\nicefrac{3}{2}\right)}}+\frac{1}{\hat{\epsilon}_{\kappa}}\right)-\frac{\left(\eta_{\kappa}-\eta_{\kappa}^{\prime}\right)^{2}}{\epsilon_{\kappa}^{\left(\nicefrac{3}{2}\right)}\hat{\epsilon}_{\kappa}}\right]=0.\label{eq:MGZ18:DE-BK-OAW}
\end{multline}

Some of the particular cases of equation (\ref{eq:MGZ18:DE-BK-OAW})
are discussed below.

\subsection{Maxwellian limit}

If one takes the limits $\kappa_{e}\to\infty$ and $\kappa_{i}\to\infty$,
one observes that $1_{\kappa}^{(\beta)}\xrightarrow{\kappa\to\infty}1$,
and the limits of the kappa plasma functions are given in Papers I
and II\@. In this case, the dispersion equation (\ref{eq:MGZ18:DE-BK-OAW}),
as well as the parameters (\ref{eq:MGZ18:DT-BK-pars}a-g) reduce to
the corresponding forms given by Ref. \onlinecite{YoonWuAssis93/07},
as expected.

\subsection{Limit of parallel propagation}

In the limit $k_{\perp}\to0$ or, equivalently, the limit of zero
ion gyroradius, we have $\mu_{i}\to0$, and we have to employ the
limiting forms given by (I.21), (II. 23), the expression given in
page 04 of Paper II, and (\ref{eq:MGZ18:dY-dxi-mu0})\@. As a result,
the dispersion equation (\ref{eq:MGZ18:DE-BK-OAW}) reduces to 
\begin{gather*}
\left(\frac{\omega^{2}}{\Omega_{i}^{2}}+\eta\frac{\omega}{\Omega_{i}}+\gamma\right)\left(\frac{\omega^{2}}{\Omega_{i}^{2}}-\eta\frac{\omega}{\Omega_{i}}+\gamma\right)=0,\\
\begin{aligned}\gamma & =\frac{1}{2}\left(\beta_{\parallel e}A_{e}+\beta_{\parallel i}A_{i}-2\right)\frac{k_{\parallel}^{2}v_{A}^{2}}{\Omega_{i}^{2}}\\
\eta & =-\frac{1}{2}\left(1+2A_{i}\right)\beta_{\parallel i}\frac{k_{\parallel}^{2}v_{A}^{2}}{\Omega_{i}^{2}},
\end{aligned}
\end{gather*}
which is exactly the same solution obtained by Ref. \onlinecite{YoonWuAssis93/07}\@. 

In this case, the dispersion relations are the roots of polynomials
that only depend on the wavenumber, the plasma betas and the anisotropy
parameters. In other words, the dispersion equation (\ref{eq:MGZ18:DE-BK-OAW})
predicts that the kappa parameters $\kappa_{e}$ and $\kappa_{i}$
do not influence the parallel firehose instability.

An exact treatment of the PFH shows a different picture. Whereas the
real part of the dispersion relation is largely independent on the
kappas, the growth rate does depend on $\kappa_{e}$ and, mostly,
on $\kappa_{i}$\@. The reason why the treatment presented here is
not able to account for this fact lies with the assumptions made about
the magnitudes of the quantities $\left|\xi_{\pm e}\right|$ and $\left|\xi_{\pm i}\right|$\@.
When $k_{\perp}=0$, the ion-firehose occurs in the right-handed mode
(the magnetosonic mode)\@. A careful examination of the unstable
range shows that the nonresonant approximations $\left|\xi_{\pm e}\right|\gg1$
and $\left|\xi_{+i}\right|\gg1$ are still valid, but $\left|\xi_{-i}\right|\simeq1$\@.
Hence, for the parallel firehose instability the kinetic effects due
to this quantity can not be ignored and the growth rate is no longer
given by the simple root of a polynomial, as above. A detailed account
of the PFH in a bi-kappa plasma is given by Ref. \onlinecite{Vinas+17}.

\subsection{Limit of perpendicular propagation}

In the converse case $\left(k_{\parallel}\to0\right)$, one must consider
in (\ref{eq:MGZ18:DT-BK-pars}a-g) the asymptotic forms of the kappa
plasma functions for the limits $\xi_{0e},\xi_{0i}\to\infty$\@.
For the function $Z_{\kappa}^{\left(\beta\right)}\left(\xi\right)$
one can notice in the expansion given by the expression at page 14
of Paper I that the dominant term is $Z_{\kappa}^{\left(\beta\right)}\left(\xi\right)\simeq-1_{\kappa}^{\left(\beta+\nicefrac{1}{2}\right)}/\xi$\@.
For the functions $\mathcal{Z}$ and $\mathcal{Y}$, one employs again
the dominant terms of expansions (II.28a,b).

After some manipulations, equation (\ref{eq:MGZ18:DE-BK-OAW}) factors
into two branches: $\omega^{2}=0$ and
\begin{align*}
\left(\frac{\omega}{\Omega_{i}}\right)^{2} & =\frac{1}{\epsilon_{\kappa}^{\left(\nicefrac{3}{2}\right)\prime}}\left(d_{3\kappa}+\frac{d_{4\kappa}^{2}}{\epsilon_{\kappa}^{\left(\nicefrac{3}{2}\right)}}\right),\text{ where}\\
d_{3\kappa} & =\left[1+\beta_{\perp e}-\beta_{\perp i,\kappa}\mathrm{H}_{0,\kappa_{i}}^{\left(\nicefrac{1}{2}\right)\prime}\left(\mu_{i}\right)\right]\frac{2\mu_{i}}{\beta_{\perp i,\kappa}}\\
d_{4\kappa} & =1_{\kappa_{e}}^{\left(\nicefrac{1}{2}\right)}+\mathrm{H}_{0,\kappa_{i}}^{\left(\nicefrac{3}{2}\right)\prime}\left(\mu_{i}\right).
\end{align*}

In the Maxwellian limit, the expression of the nonzero mode, in the
lowest-order of a small gyroradius expansion, is
\[
\omega\simeq\sqrt{1+\beta_{\perp e}+\beta_{\perp i}}k_{\perp}v_{A},
\]
which corresponds to the dispersion relation of compressive Alfvén
waves. These waves are always damped. On the other hand, the nonpropagating
mode with $\omega^{2}=0$ turns out to be the unstable mode.

For any combination of $\left(k_{\perp},k_{\parallel}\right)$, the
wave modes are obtained from the numerical solution of equation (\ref{eq:MGZ18:DE-BK-OAW})\@.
Several new properties of the kappa plasma functions, not included
in Papers I and II, were needed in order to implement the numerical
solution of the dispersion equation. These properties are discussed
in section \ref{sec:MGZ18:Kappa_plasma_functions}, and some numerical
solutions are presented in section \ref{sec:MGZ18:Numerical_sols}.

\section{New expressions for the kappa plasma special functions\label{sec:MGZ18:Kappa_plasma_functions}}

Several new properties and representations for the kappa special functions
are developed here.

\subsection{Symmetry properties of the superthermal plasma dispersion function}

The superthermal (or kappa) plasma dispersion function ($\kappa$PDF)
$Z_{\kappa}^{\left(\alpha,\beta\right)}\left(\xi\right)$ was initially
defined in (I.11) and several properties were presented in Papers
I and II\@. Here, we will show its symmetry properties, which will
be important for the new expansions of the functions $\mathcal{Z}$
and $\mathcal{Y}$ to be presented below.

Consider first the argument $\xi=\xi_{r}+i\xi_{i}\in\mathbb{C}$\@.
If $\xi_{i}>0$, one of the possible representations of $Z_{\kappa}^{\left(\alpha,\beta\right)}\left(\xi\right)$
is given by (I.13), while its analytic continuation for $\xi_{i}\leqslant0$
is given by (I.A1)\@. Hence, we can directly establish the first
identity, when $\xi_{i}>0$,\begin{subequations}\label{eq:MGZ18:PDF-Symmetry-props}
\begin{multline}
Z_{\kappa}^{\left(\alpha,\beta\right)}\left(-\xi\right)+Z_{\kappa}^{\left(\alpha,\beta\right)}\left(\xi\right)\\
=2i\sqrt{\pi}\frac{\kappa^{-\beta-1/2}\Gamma\left(\lambda-1\right)}{\Gamma\left(\sigma-\nicefrac{3}{2}\right)}\left(1+\frac{\xi^{2}}{\kappa}\right)^{-\left(\lambda-1\right)}.\label{eq:MGZ18:PDF-Symmetry-1}
\end{multline}

As discussed in Papers I and II, when $\alpha=\beta=1$, the function
$Z_{\kappa}^{\left(1,1\right)}\left(\xi\right)\equiv Z_{\kappa}\left(\xi\right)$
reduces to the dispersion function first introduced by Summers and
Thorne\citep{SummersThorne91/08} and largely discussed by Mace and
Hellberg\citep{MaceHellberg95/06}\@. In this case, the symmetry
property above reduces to Eq. 27 of Ref. \onlinecite{MaceHellberg95/06}\@.
In the Maxwellian limit $\left(\kappa\to\infty\right)$, the same
property reduces to the known identity for the Fried \& Conte function,\citep{FriedConte61}
\[
Z\left(-\xi\right)+Z\left(\xi\right)=2i\sqrt{\pi}e^{-\xi^{2}},\:\left(\xi_{i}>0\right).
\]

Now, let us consider the representation (I.14), which is valid in
the principal branch $-\frac{1}{2}\pi<\arg\left(\xi+i\sqrt{\kappa}\right)<\frac{3}{2}\pi$
(\emph{i.e.,} the branch line runs along $-i\sqrt{\kappa}<\xi<-i\infty$)\@.
This is the generalization of the result given by Eq. 17 of Ref. \onlinecite{MaceHellberg95/06}\@.
If we denote by $\bar{\xi}=\xi_{r}-i\xi_{i}$ the operation of complex
conjugation, we immediately observe that 
\begin{align*}
Z_{\kappa}^{\left(\alpha,\beta\right)}\left(-\bar{\xi}\right) & =i\frac{\kappa^{-\beta-1/2}\Gamma\left(\lambda-\nicefrac{1}{2}\right)}{\left(\lambda-1\right)\Gamma\left(\sigma-\nicefrac{3}{2}\right)}\\
 & \times\pFq 21\left[{1,2\left(\lambda-1\right)\atop \lambda};\frac{1}{2}\left(1-\frac{i\bar{\xi}}{\kappa^{1/2}}\right)\right].
\end{align*}
However, given that\footnote{See, \emph{e.g., }\url{http://functions.wolfram.com/07.23.04.0003.02}.}
\[
\pFq 21\left({\bar{a},\bar{b}\atop \bar{c}};\bar{z}\right)=\overline{\pFq 21\left({a,b\atop c};z\right)}\quad\left(z\notin\left(1,\infty\right)\right),
\]
which is a property shared by all hypergeometric functions in their
principal branches, it follows that
\begin{equation}
Z_{\kappa}^{\left(\alpha,\beta\right)}\left(\bar{\xi}\right)=-\overline{Z_{\kappa}^{\left(\alpha,\beta\right)}\left(-\xi\right)}.\label{eq:MGZ18:PDF-Symmetry-2}
\end{equation}
This result also reduces to the known property $Z\left(\bar{\xi}\right)=-\overline{Z\left(-\xi\right)}$.

Finally, by combining properties (\ref{eq:MGZ18:PDF-Symmetry-1})
and (\ref{eq:MGZ18:PDF-Symmetry-2}), we obtain, for $\xi_{i}>0$,
\begin{equation}
\begin{aligned}Z_{\kappa}^{\left(\alpha,\beta\right)}\left(\bar{\xi}\right) & =\overline{Z_{\kappa}^{\left(\alpha,\beta\right)}\left(\xi\right)}\\
 & +2i\sqrt{\pi}\frac{\kappa^{-\beta-1/2}\Gamma\left(\lambda-1\right)}{\Gamma\left(\sigma-\nicefrac{3}{2}\right)}\left(1+\frac{\bar{\xi}^{2}}{\kappa}\right)^{-\left(\lambda-1\right)},
\end{aligned}
\label{eq:MGZ18:PDF-Symmetry-3}
\end{equation}
which, when $\kappa\to\infty$, reduces to 
\[
Z\left(\bar{\xi}\right)=\overline{Z\left(\xi\right)}+2i\sqrt{\pi}e^{-\bar{\xi}^{2}}.
\]

\end{subequations}

\subsection{The superthermal plasma gyroradius function}

The kappa plasma gyroradius function ($\kappa$PGF) $\mathcal{H}_{n,\kappa}^{\left(\alpha,\beta\right)}\left(z\right)$
was initially defined by (I.20) and its more general representation
was given by (I.22) in terms of the Meijer $G$-function. The definition
and properties of the $G$-function can be seen in appendix B of Paper
I or in the references given therein. Several properties of $\mathcal{H}_{n,\kappa}^{\left(\alpha,\beta\right)}\left(z\right)$
and an associated function are discussed in Papers I and II\@. Here
we will present some new properties of the $\kappa$PGF.

\subsubsection{Symmetry properties}

In some of the representations shown in Papers I and II as well as
in the present work, the function $\mathcal{H}_{n,\kappa}^{\left(\alpha,\beta\right)}\left(z\right)$
needs to be evaluated for complex $z$\@. Looking at the representations
(I.23, 24), respectively valid for noninteger/integer values of $\lambda=\kappa+\alpha+\beta$,
and taking into account the symmetry properties of elementary and
Bessel functions, one readily concludes \emph{(i)} that the $\kappa$PGF
is a multivalued function with the origin as a branch point and with
a branch cut along the negative real axis, and\emph{ (ii)} that in
the principal branch,

\begin{equation}
\mathcal{H}_{n,\kappa}^{\left(\alpha,\beta\right)}\left(\bar{z}\right)=\overline{\mathcal{H}_{n,\kappa}^{\left(\alpha,\beta\right)}\left(z\right)},\qquad\left(\left|\arg z\right|<\pi\right).\label{eq:MGZ18:PGF-symmetry}
\end{equation}

\subsubsection{Series expansion in the integer case}

As it was argued at length in Papers I and II, the function $\mathcal{H}_{n,\kappa}^{\left(\alpha,\beta\right)}\left(z\right)$
can be represented in terms of power series only when $\lambda=\kappa+\alpha+\beta$
is not integer. When $\lambda\in\mathbb{Z}$, the function has a logarithmic
term and a representation in terms of modified Bessel functions was
given by (I.24) (see Eq. II.7c for the derivatives)\@. Although general,
the evaluation of $\mathcal{H}_{n,\kappa}^{\left(\alpha,\beta\right)}\left(z\right)$
via the Bessel functions can be computationally expensive when $\left|z\right|<1$,
as is the case for the application discussed in this paper. Hence,
a suitable series expansion for $\mathcal{H}_{n,\kappa}^{\left(\alpha,\beta\right)}\left(z\right)$
in the regime of small gyroradius is desired.

The desired representation is obtained via the residue theorem. Using
(I.22) and (I.B10), the function $\mathcal{H}_{n,\kappa}^{\left(\alpha,\beta\right)}\left(z\right)$
can be written as\begin{subequations}\label{eq:MGZ18:H-integer}
\begin{align}
\mathcal{H}_{n,\kappa}^{\left(\alpha,\beta\right)}\left(z\right) & =\frac{1}{\sqrt{\pi}}\frac{\kappa}{\Gamma\left(\lambda-1\right)}F_{n,\lambda-2}\left(2\kappa z\right),\label{eq:MGZ18:H_n,k^a,b-Integer}\\
F_{n,\theta}\left(z\right) & =\frac{1}{2\pi i}\int_{L}f_{n,\theta}\left(s\right)z^{-s}ds,\nonumber \\
f_{n,\theta}\left(s\right) & =\frac{\Gamma\left(\theta+s\right)\Gamma\left(n+s\right)\Gamma\left(\nicefrac{1}{2}-s\right)}{\Gamma\left(n+1-s\right)},\nonumber 
\end{align}
where the integration contour $L$ is such that all poles of $\Gamma\left(\theta+s\right)$
and $\Gamma\left(n+s\right)$ lie to the right of $L$ whereas the
poles of $\Gamma\left(\nicefrac{1}{2}-s\right)$ lie to the left.
In this case, the residues of the integration are given only by the
poles of the first two gamma functions.

If $m\in\mathbb{Z}$, when $\theta=m$ (for $\mathcal{H}_{n,\kappa}^{\left(\alpha,\beta\right)}$,
$m=0,1,2,\dots$), the poles of $f_{n,\theta}\left(s\right)$ occur
at $s=-m-r$ and $s=-n-r$ $\left(r\in\mathbb{N}\right)$\@. Hence,
there are two possibilities: \emph{(i) }$m<n$ and \emph{(ii)} $m\geqslant n$,
that will be treated separately.

\paragraph{1\@. $\boldsymbol{m<n}$:}

the function $f_{n,\theta}\left(s\right)$ has simple poles at $s=-m,-m-1,\dots,-n+1$
and double poles at $s=-n,-n-1,\dots$\@. The residues of the simple
poles can be evaluated with the identity\footnote{See \url{http://functions.wolfram.com/06.05.06.0007.01}.}
\[
\Gamma\left(z\right)=\frac{\left(-\right)^{k}}{k!\left(z+k\right)}+\frac{\left(-\right)^{k}}{k!}\psi\left(k+1\right)+\mathcal{O}\left[\left(z+k\right)\right],
\]
where $\psi\left(z\right)$ is the psi or digamma function. This expansion
is valid for $z\simeq-k$ $\left(k=0,1,2,\dots\right)$\@. Therefore,
\[
\mathrm{Res}f_{n,\theta}\left(s=-\ell\right)=\sqrt{\pi}\left(-\right)^{m}\frac{\Gamma\left(n-\ell\right)\left(\nicefrac{1}{2}\right)_{\ell}}{\left(1\right)_{n+\ell}\left(1\right)_{\ell-m}}\left(-z\right)^{\ell},
\]
where $\ell=m,m+1,\dots,n-1$ and $\left(a\right)_{r}=\Gamma\left(a+r\right)/\Gamma\left(a\right)$
is the Pochhammer symbol.

On the other hand, the double poles are obtained from
\[
\mathrm{Res}f_{n,\theta}\left(s=-\ell\right)=\lim_{s\to-\ell}\frac{d}{ds}\left[\left(s+\ell\right)^{2}f_{n,\theta}\left(s\right)\right],
\]
for $\ell=n,n+1,\dots$\@. Given that $\Gamma^{\prime}\left(z\right)=\psi\left(z\right)\Gamma\left(z\right)$,
and using the identity\footnote{See \url{http://functions.wolfram.com/06.14.06.0010.02}.}
\[
\psi\left(z\right)=-\frac{1}{z+k}+\psi\left(k+1\right)+\mathcal{O}\left[\left(z+k\right)\right],
\]
one obtains for the residues
\begin{align*}
\mathrm{Res}f_{n,\theta}\left(s=-\ell\right) & =\frac{\sqrt{\pi}\left(\nicefrac{1}{2}\right)_{\ell}\left(-\right)^{m+n}z^{\ell}}{\left(1\right)_{n+\ell}\left(1\right)_{\ell-m}\left(1\right)_{\ell-n}}\\
 & \times\left[\vphantom{\frac{1}{2}}\psi\left(\ell-m+1\right)+\psi\left(\ell-n+1\right)\right.\\
 & \left.-\psi\left(\frac{1}{2}+\ell\right)+\psi\left(n+\ell+1\right)-\ln z\right].
\end{align*}

\paragraph{2\@. $\boldsymbol{m\geqslant n}$:}

now the poles are simple for $s=-n,\dots,-m+1$ $\left(m>n\right)$
and double for $s=-m,-m-1,\dots$\@. The evaluation of the residues
follows the same lines as in the previous case.

Summing the contributions from all residues, both possibilities can
be cast in a single expression for $F_{n,\theta}\left(z\right)$,
resulting finally
\begin{equation}
\begin{aligned}\frac{F_{n,\theta}\left(z\right)}{\sqrt{\pi}} & =\epsilon_{\sigma}\left(\sigma-1\right)!\left(\frac{1}{2}\right)_{\nu}\frac{z^{\nu}}{\eta!}\sum_{\ell=0}^{\sigma-1}\frac{\left(\nu+\nicefrac{1}{2}\right)_{\ell}}{\left(\eta+1\right)_{\ell}\left(1-\sigma\right)_{\ell}}\frac{z^{\ell}}{\ell!}\\
 & -\left(-\right)^{\nu}\left(\frac{1}{2}\right)_{\mu}\frac{\left(-z\right)^{\mu}}{\psi!\sigma!}\\
 & \times\left[\left(2\gamma+\ln z\right)\pFq 12\left({\mu+\nicefrac{1}{2}\atop \psi+1,\sigma+1};z\right)\right.\\
 & -\sum_{\ell=0}^{\infty}\frac{\left(\mu+\nicefrac{1}{2}\right)_{\ell}}{\left(\psi+1\right)_{\ell}\left(\sigma+1\right)_{\ell}}\\
 & \left.\times\left(H_{\ell}+H_{\psi+\ell}+H_{\sigma+\ell}-H_{\mu-\nicefrac{1}{2}+\ell}\right)\frac{z^{\ell}}{\ell!}\right],
\end{aligned}
\label{eq:MGZ18:Fng}
\end{equation}
where $\mu=\max\left(\theta,n\right)$, $\nu=\min\left(\theta,n\right)$,
$\sigma=\mu-\nu$, $\psi=\mu+n$, $\eta=\nu+n$, and $\epsilon_{\sigma}=1-\delta_{\sigma,0}$\@.
In (\ref{eq:MGZ18:Fng}), the quantities $H_{w}$ are the harmonic
numbers, given by $H_{w}=\psi\left(w+1\right)+\gamma$, where $\gamma=0.57721\dots$
is Euler's constant. When $w$ is a nonnegative integer, the harmonic
numbers are $H_{0}=0$ and $H_{\ell}=\sum_{k=1}^{\ell}k^{-1}$.

\end{subequations}

Notice that there are two terms in (\ref{eq:MGZ18:Fng}) that distinguishes
the series expansion of $\mathcal{H}_{n,\kappa}^{\left(\alpha,\beta\right)}\left(z\right)$
when $\lambda$ is integer from a simple power series expansion of
the hypergeometric kind. First, there is a term proportional to $\ln z$
and second, the last term is a power series but is not hypergeometric.

\subsection{The two-variables kappa plasma functions}

The two-varibles kappa plasma functions (2VKPs) $\mathcal{Z}_{n,\kappa}^{\left(\alpha,\beta\right)}\left(\mu,\xi\right)$
and $\mathcal{Y}_{n,\kappa}^{\left(\alpha,\beta\right)}\left(\mu,\xi\right)$
were initially defined by (I.26)\@. They describe the dispersive
properties of oscillations occuring in a magnetized (bi-)kappa plasma.
Since a magnetized plasma described by a $\kappa$VDF displays strong
correlations between the parallel and perpendicular components of
the particles' velocities, the 2VKPs can not be factored as the product
of two simple functions; \emph{i.e.,} there are no functions $M\left(\mu\right)$
and $N\left(\xi\right)$ such that $\mathcal{Z}\left(\mu,\xi\right)=M\left(\mu\right)N\left(\xi\right)$,
for instance. This point has been argued at length in Ref. \onlinecite{GaelzerZiebell14/12}
and in Papers I and II\@. Consequently, the simpler representations
one can expect for these functions will always be some transcendental
series expansion.

\subsubsection{Symmetry properties}

The symmetry properties of the functions $\mathcal{Z}_{n,\kappa}^{\left(\alpha,\beta\right)}\left(\mu,\xi\right)$
and $\mathcal{Y}_{n,\kappa}^{\left(\alpha,\beta\right)}\left(\mu,\xi\right)$
can be readily derived starting from the integral representations
(I.26a,b)\@. For applications to plasma physics, $\mu>0$ always
but $\xi$ is in general complex. Hence, the symmetry properties of
the 2VKPs are determined by the $Z_{\kappa}^{\left(\alpha,\beta\right)}\left(\xi\right)$
function.

Consequently, from the symmetry properties (\ref{eq:MGZ18:PDF-Symmetry-props}),
one concludes that\begin{subequations}\label{eq:MGZ18:ZY-symmetry-props}
\begin{align}
\mathcal{Z}_{n,\kappa}^{\left(\alpha,\beta\right)}\left(\mu,\bar{\xi}\right) & =-\overline{\mathcal{Z}_{n,\kappa}^{\left(\alpha,\beta\right)}\left(\mu,-\xi\right)}\label{eq:MGZ18:Z-symmetry-1}\\
\mathcal{Y}_{n,\kappa}^{\left(\alpha,\beta\right)}\left(\mu,\bar{\xi}\right) & =-\overline{\mathcal{Y}_{n,\kappa}^{\left(\alpha,\beta\right)}\left(\mu,-\xi\right)}\label{eq:MGZ18:Y-symmetry-1}\\
\mathcal{Z}_{n,\kappa}^{\left(\alpha,\beta\right)}\left(\mu,\bar{\xi}\right) & =\overline{\mathcal{Z}_{n,\kappa}^{\left(\alpha,\beta\right)}\left(\mu,\xi\right)}\nonumber \\
 & +2i\sqrt{\pi}\frac{\kappa^{-1/2-\beta}\Gamma\left(\lambda-1\right)}{\Gamma\left(\sigma-\nicefrac{3}{2}\right)}\left(1+\frac{\bar{\xi}^{2}}{\kappa}\right)^{-\left(\lambda-2\right)}\nonumber \\
 & \times\mathcal{H}_{n,\kappa}^{\left(\alpha,\beta\right)}\left[\mu\left(1+\frac{\bar{\xi}^{2}}{\kappa}\right)\right]\label{eq:MGZ18:Z-symmetry-2}\\
\mathcal{Y}_{n,\kappa}^{\left(\alpha,\beta\right)}\left(\mu,\bar{\xi}\right) & =\overline{\mathcal{Y}_{n,\kappa}^{\left(\alpha,\beta\right)}\left(\mu,\xi\right)}\nonumber \\
 & +2i\sqrt{\pi}\frac{\kappa^{1/2-\beta}\Gamma\left(\lambda-2\right)}{\Gamma\left(\sigma-\nicefrac{3}{2}\right)}\left(1+\frac{\bar{\xi}^{2}}{\kappa}\right)^{-\left(\lambda-4\right)}\nonumber \\
 & \times\mathcal{H}_{n,\kappa}^{\left(\alpha,\beta-1\right)\prime}\left[\mu\left(1+\frac{\bar{\xi}^{2}}{\kappa}\right)\right].\label{eq:MGZ18:Y-symmetry-2}
\end{align}
For the derivation of (\ref{eq:MGZ18:Z-symmetry-2}) and (\ref{eq:MGZ18:Y-symmetry-2})
we have employed the identities (II.24, 26).

Moreover, it can be easily verified that 
\begin{equation}
\begin{aligned}\mathcal{Z}_{-n,\kappa}^{\left(\alpha,\beta\right)}\left(\mu,\xi\right) & =\mathcal{Z}_{n,\kappa}^{\left(\alpha,\beta\right)}\left(\mu,\xi\right),\\
\mathcal{Y}_{-n,\kappa}^{\left(\alpha,\beta\right)}\left(\mu,\xi\right) & =\mathcal{Y}_{n,\kappa}^{\left(\alpha,\beta\right)}\left(\mu,\xi\right).
\end{aligned}
\label{eq:MGZ18:XY-n-symmetry}
\end{equation}
Henceforth, we will implicitly assume that $n\geqslant0$.\end{subequations}

\subsubsection{Derivative of $\boldsymbol{\mathcal{Y}_{n,\kappa}^{\left(\alpha,\beta\right)}\left(\mu,\xi\right)}$}

As can be observed in the expressions for the dielectric tensor given
either by (II.3) or by (\ref{eq:MGZ18:DT-BK-HB}) and (\ref{eq:MGZ18:Coefs-ABC}),
one also needs to evaluate $\partial_{\xi}\mathcal{Y}_{n,\kappa}^{\left(\alpha,\beta\right)}\left(\mu,\xi\right)$\@.
The expressions for this derivative were not included in Paper II
due to an oversight, which will be remedied now.

The procedure is roughly the same as the one described in section
C.1 of Paper II for the function $\mathcal{Z}$\@. So, without further
ado, we present 
\begin{align}
\partial_{\xi}\mathcal{Y}_{n,\kappa}^{\left(\alpha,\beta\right)}\left(\mu,\xi\right) & =-2\left[\frac{\Gamma\left(\lambda-\nicefrac{3}{2}\right)}{\kappa^{\beta}\Gamma\left(\sigma-\nicefrac{3}{2}\right)}\mathcal{H}_{n,\kappa}^{\left(\alpha,\beta-\nicefrac{1}{2}\right)\prime}\left(\mu\right)\right.\nonumber \\
 & \left.+\xi\mathcal{Y}_{n,\kappa}^{\left(\alpha,\beta+1\right)}\left(\mu,\xi\right)\vphantom{\frac{\Gamma\left(\lambda-\nicefrac{3}{2}\right)}{\Gamma\left(\sigma-\nicefrac{3}{2}\right)}}\right],\label{eq:MGZ18:dY_cal-dxi}
\end{align}
with the particular case,
\begin{equation}
\partial_{\xi}\mathcal{Y}_{n,\kappa}^{\left(\alpha,\beta\right)}\left(0,\xi\right)=\left(-\delta_{n,0}+\frac{1}{2}\delta_{\left|n\right|,1}\right)Z_{\kappa}^{\left(\alpha,\beta-3\right)\prime}\left(\xi\right).\label{eq:MGZ18:dY-dxi-mu0}
\end{equation}

\subsubsection{Power series expansions for large $\boldsymbol{\left|\xi\right|}$}

Power series expansions for $\mathcal{Z}$ and $\mathcal{Y}$ were
given by (II.25a) and (II.27a), which are formally convergent for
$\left|\xi\right|<\sqrt{\kappa}$\@. Additionally, other series expansions
were given by (II.25c) and (II.27b), which are formally valid for
the whole complex plane of $\xi$\@. However, for the range of variations
of the physical parameters considered in this work, although the argument
$\mu$ is always small, since it is proportional to the particle's
gyroradius, the magnitude of the argument $\xi=\xi_{0s}=\omega/k_{\parallel}w_{\parallel s}$
can vary from very small (for near-parallel propagation) to very large
(for near-perpendicular propagation)\@. Hence, from the computational
point of view it is desirable to have power series expansions for
the 2VKPs valid for $\left|\xi\right|>\sqrt{\kappa}$\@. These series
will now be derived.\\
\begin{widetext}

We will consider first the function $\mathcal{Z}_{n,\kappa}^{\left(\alpha,\beta\right)}\left(\mu,\xi\right)$\@.
Taking the integral representation (I.26a) and inserting representation
(II.15a) for $Z_{\kappa}^{\left(\alpha,\beta\right)}\left(\xi\right)$,
we can integrate the last term using identity (II.24) and obtain
\begin{align*}
\mathcal{Z}_{n,\kappa}^{\left(\alpha,\beta\right)}\left(\mu,\xi\right) & =-2\frac{\pi^{1/2}\kappa^{-\beta-1/2}}{\Gamma\left(\sigma-\nicefrac{3}{2}\right)}\frac{\xi}{\sqrt{\kappa}}F_{2}\left(\mu,\xi\right)+\frac{i\pi^{1/2}\Gamma\left(\lambda-1\right)}{\kappa^{\beta+1/2}\Gamma\left(\sigma-\nicefrac{3}{2}\right)}\left(1+\frac{\xi^{2}}{\kappa}\right)^{-\left(\lambda-2\right)}\mathcal{H}_{n,\kappa}^{\left(\alpha,\beta\right)}\left[\mu\left(1+\frac{\xi^{2}}{\kappa}\right)\right],\\
F_{2}\left(\mu,\xi\right) & =\int_{0}^{\infty}dx\,\frac{xJ_{n}^{2}\left(\sqrt{2\mu}\,x\right)}{\left(1+x^{2}/\kappa\right)^{\lambda-1/2}}G_{2,2}^{1,2}\left[\frac{\xi^{2}/\kappa}{1+x^{2}/\kappa}\left|{0,\nicefrac{3}{2}-\lambda\atop 0,-\nicefrac{1}{2}}\right.\right],
\end{align*}
where $G_{2,2}^{1,2}$ is the Meijer $G$-function defined by (I.B10)\@.
Using the Mellin-Barnes representation for this function and interchanging
the integrations in $F_{2}\left(\mu,\xi\right)$, we can integrate
on $x$, resulting then 
\[
F_{2}\left(\mu,\xi\right)=\frac{\kappa}{2\sqrt{\pi}}\frac{\kappa}{\xi^{2}}\frac{1}{2\pi i}\int_{L_{t}}dt\,\frac{\Gamma\left(n-t\right)\Gamma\left(\nicefrac{1}{2}+t\right)}{\Gamma\left(n+1+t\right)}\left(2\kappa\mu\right)^{t}G_{2,2}^{2,1}\left[\frac{\kappa}{\xi^{2}}\left|{0,\nicefrac{1}{2}\atop \lambda-\nicefrac{5}{2}-t,0}\right.\right],
\]
where we have also employed the identity (I.B11a).

Let us now define the auxiliary functions 
\begin{align*}
h_{1}\left(x,y\right) & =\frac{1}{2\pi i}\int_{L_{t}}dt\,\frac{\Gamma\left(n-t\right)\Gamma\left(\nicefrac{1}{2}+t\right)}{\Gamma\left(n+1+t\right)}x^{t}G_{2,2}^{2,1}\left[y\left|{0,\nicefrac{1}{2}\atop \lambda-\nicefrac{5}{2}-t,0}\right.\right]\\
h_{2}\left(x,y\right) & =\frac{1}{2\pi^{2}i}\int_{L_{t}}dt\,\frac{\Gamma\left(\nicefrac{7}{2}-\lambda+t\right)\Gamma\left(\lambda-\nicefrac{5}{2}-t\right)\Gamma\left(n-t\right)\Gamma\left(\nicefrac{1}{2}+t\right)}{\Gamma\left(n+1+t\right)}x^{t}G_{2,2}^{1,2}\left[y\left|{0,\nicefrac{1}{2}\atop 0,\lambda-\nicefrac{5}{2}-t}\right.\right].
\end{align*}
Obviously, $F_{2}\left(\mu,\xi\right)=h_{1}\left(2\kappa\mu,\kappa/\xi^{2}\right)$\@.
Noticing that the integrations in both functions are defined along
the same integration contour, we will now evaluate $h_{1}-h_{2}$\@.
Introducing again the Mellin-Barnes representations for the $G$-functions
and simplifying the resulting expression with the help of properties
of the gamma function, we obtain, after some amount of algebra, 
\begin{align}
h_{1}-h_{2} & =-\sqrt{\pi}\frac{\Gamma\left(\lambda-1\right)}{\kappa}y^{\lambda-\nicefrac{5}{2}}\left(1+y\right)^{-\left(\lambda-2\right)}\mathcal{I}_{n,\kappa}^{\left(\alpha,\beta\right)}\left[\frac{x}{2\kappa}\left(1+\frac{1}{y}\right)\right],\text{ where}\nonumber \\
\mathcal{I}_{n,\kappa}^{\left(\alpha,\beta\right)}\left(z\right) & =\frac{1}{\sqrt{\pi}}\frac{\kappa}{\Gamma\left(\lambda-1\right)}G_{2,4}^{2,2}\left[2\kappa z\left|{\lambda-\nicefrac{5}{2},\nicefrac{1}{2}\atop \lambda-\nicefrac{5}{2},n,-n,\lambda-2}\right.\right].\label{eq:MGZ18:I_n,k^a,b}
\end{align}
In order to obtain this result, we used representations (I.B9, B10
and B14), which give 
\[
G_{1,1}^{1,1}\left[z\left|{\nicefrac{1}{2}\atop \lambda-\nicefrac{5}{2}-t}\right.\right]=\Gamma\left(\lambda-2-t\right)z^{\lambda-\nicefrac{5}{2}-t}\pFq 10\left({\lambda-2-t\atop -};-z\right)=\Gamma\left(\lambda-2-t\right)z^{\lambda-\nicefrac{5}{2}-t}\left(1+z\right)^{-\left(\lambda-2-t\right)}.
\]

Now, using again (I.B14), we can write $h_{2}\left(x,y\right)$ as
\[
h_{2}\left(x,y\right)=\frac{1}{\sqrt{\pi}}\frac{1}{2\pi i}\int_{L_{t}}dt\,\frac{\Gamma\left(\lambda-\nicefrac{5}{2}-t\right)\Gamma\left(n-t\right)\Gamma\left(\nicefrac{1}{2}+t\right)}{\Gamma\left(n+1+t\right)}x^{t}\pFq 21\left({1,\nicefrac{1}{2}\atop \nicefrac{7}{2}-\lambda+t};-y\right).
\]
Noticing that in $F_{2}\left(\mu,\xi\right)$ there is now a term
with $h_{2}\left(2\kappa\mu,\kappa/\xi^{2}\right)$ and that we are
assuming that $\left|\xi\right|>\sqrt{\kappa}$, we can then formally
expand the Gauss function in $h_{2}$ according to (I.B4) and identify
the remaining integral as a $G$-function.

Therefore, we arrive at the desired result,\begin{subequations}\label{eq:MGZ18:XY-asymp-xi}
\begin{align}
\mathcal{Z}_{n,\kappa}^{\left(\alpha,\beta\right)}\left(\mu,\xi\right) & =\frac{\sqrt{\pi}\Gamma\left(\lambda-1\right)}{\kappa^{1/2+\beta}\Gamma\left(\sigma-\nicefrac{3}{2}\right)}\left(1+\frac{\xi^{2}}{\kappa}\right)^{-\left(\lambda-2\right)}\left\{ \mathcal{I}_{n,\kappa}^{\left(\alpha,\beta\right)}\left[\mu\left(1+\frac{\xi^{2}}{\kappa}\right)\right]+i\mathcal{H}_{n,\kappa}^{\left(\alpha,\beta\right)}\left[\mu\left(1+\frac{\xi^{2}}{\kappa}\right)\right]\right\} \nonumber \\
 & -\frac{\kappa^{-1/2-\beta}}{\Gamma\left(\sigma-\nicefrac{3}{2}\right)}\frac{\sqrt{\kappa}}{\xi}\mathcal{X}_{n,\kappa}^{\left(\alpha,\beta\right)}\left(\mu,\frac{\kappa}{\xi^{2}}\right),\label{eq:MGZ18:X-asymp-xi}\\
\mathcal{X}_{n,\kappa}^{\left(\alpha,\beta\right)}\left(x,y\right) & =\sum_{k=0}^{\infty}\left(\frac{1}{2}\right)_{k}\mathcal{J}_{n,k,\kappa}^{\left(\alpha,\beta\right)}\left(x\right)y^{k},\nonumber \\
\mathcal{J}_{n,k,\kappa}^{\left(\alpha,\beta\right)}\left(x\right) & =\frac{\left(-\right)^{k}\kappa}{\sqrt{\pi}}G_{2,4}^{2,2}\left[2\kappa x\left|{\lambda-\nicefrac{5}{2},\nicefrac{1}{2}\atop \lambda-\nicefrac{5}{2},n,-n,\lambda-\nicefrac{5}{2}-k}\right.\right].\nonumber 
\end{align}

Now for the function $\mathcal{Y}_{n,\kappa}^{\left(\alpha,\beta\right)}\left(\mu,\xi\right)$\@.
Starting from the integral representation (I.26b), using (II.15a,
26) and proceeding in the same manner as above, we obtain the power
series expansion 
\begin{equation}
\begin{aligned}\mathcal{Y}_{n,\kappa}^{\left(\alpha,\beta\right)}\left(\mu,\xi\right) & =\frac{\sqrt{\pi}\Gamma\left(\lambda-2\right)}{\kappa^{\beta-1/2}\Gamma\left(\sigma-\nicefrac{3}{2}\right)}\left(1+\frac{\xi^{2}}{\kappa}\right)^{-\left(\lambda-4\right)}\left\{ \mathcal{I}_{n,\kappa}^{\left(\alpha,\beta-1\right)\prime}\left[\mu\left(1+\frac{\xi^{2}}{\kappa}\right)\right]+i\mathcal{H}_{n,\kappa}^{\left(\alpha,\beta-1\right)\prime}\left[\mu\left(1+\frac{\xi^{2}}{\kappa}\right)\right]\right\} \\
 & -\frac{\kappa^{1/2-\beta}}{\Gamma\left(\sigma-\nicefrac{3}{2}\right)}\frac{\sqrt{\kappa}}{\xi}\partial_{\mu}\mathcal{X}_{n,\kappa}^{\left(\alpha,\beta-1\right)}\left(\mu,\frac{\kappa}{\xi^{2}}\right),
\end{aligned}
\label{eq:MGZ18:Y-asymp-xi}
\end{equation}
where $\partial_{x}\mathcal{X}\left(x,y\right)\equiv\partial\mathcal{X}/\partial x$.

\end{subequations}

The power series (\ref{eq:MGZ18:X-asymp-xi}) and (\ref{eq:MGZ18:Y-asymp-xi})
are formally valid for $\left|\xi\right|>\sqrt{\kappa}$\@. However,
there is an additional condition. Since the argument of $\mathcal{H}_{n,\kappa}^{\left(\alpha,\beta\right)}\left(z\right)$
is $z=\mu\left(1+\xi^{2}/\kappa\right)\in\mathbb{C}$, and since the
semiaxis $z<0$ is a branch line, we must verify when this line can
be crossed. This can happen when $\xi_{r}\to0$ for a fixed $\xi_{i}$
in the region outside the hyperbolas $\xi_{i}^{2}-\xi_{r}^{2}>\kappa$\@.
The same situation applies for the function $\mathcal{I}_{n,\kappa}^{\left(\alpha,\beta\right)}\left(z\right)$\@. 

Consequently, we can impose for the formulas (\ref{eq:MGZ18:XY-asymp-xi}a,b)
the additional validity condition $\xi_{r}\geqslant0$\@. This is
not an hindrance, however, since the analytic continuations of the
expansions (\ref{eq:MGZ18:XY-asymp-xi}a,b) for the region $\xi_{r}<0$
are evaluated with the symmetry properties (\ref{eq:MGZ18:ZY-symmetry-props}a-d).

\subsubsection{Representations for the functions $\boldsymbol{\mathcal{I}_{n,\kappa}^{\left(\alpha,\beta\right)}\left(z\right)}$
and $\boldsymbol{\mathcal{J}_{n,\kappa}^{\left(\alpha,\beta\right)}\left(z\right)}$}

The series expansions (\ref{eq:MGZ18:XY-asymp-xi}a,b) for large $\left|\xi\right|$
introduced the new associated functions $\mathcal{I}_{n,\kappa}^{\left(\alpha,\beta\right)}\left(z\right)$
and $\mathcal{J}_{n,\kappa}^{\left(\alpha,\beta\right)}\left(z\right)$,
which require adequate representations for their numerical evaluation.
These representations are derived below.

\paragraph{Function $\mathcal{I}_{n,\kappa}^{\left(\alpha,\beta\right)}\left(z\right)$\@.}

First of all, we observe in (\ref{eq:MGZ18:I_n,k^a,b}) that the function
$\mathcal{I}_{n,\kappa}^{\left(\alpha,\beta\right)}\left(z\right)$
is not defined when $\lambda$ is half-integer (with $\lambda=\nicefrac{5}{2},\nicefrac{7}{2},\dots$),
and $\lambda-\nicefrac{5}{2}-n=1,2,\dots$\@.\footnote{Recall the definition of the $G$-function given in the appendix B
of Paper I.} In this case, one can either go back to the definition of the $G$-function
and manipulate the gamma functions or employ the identity\citep{Luke69a,Prudnikov90v3}\begin{subequations}\label{eq:MGZ18:I_n,k^a,b-reps}
\[
G_{p,q+2}^{m+1,n}\left[z\left|{\left(a_{p}\right)\atop \beta,\left(b_{q}\right),\beta\pm\ell}\right.\right]=\left(-\right)^{\ell}G_{p,q+2}^{m+1,n}\left[z\left|{\left(a_{p}\right)\atop \beta\pm\ell,\left(b_{q}\right),\beta}\right.\right],\;\left(m\leqslant q\right),
\]
followed by (I.B11a)\@. Proceeding in this way, we can write
\begin{equation}
\mathcal{I}_{n,\kappa}^{\left(\alpha,\beta\right)}\left(z\right)=\frac{1}{\sqrt{\pi}}\frac{\kappa}{\Gamma\left(\lambda-1\right)}I_{n,\kappa}^{\left(\alpha,\beta\right)}\left(z\right),\text{ where }I_{n,\kappa}^{\left(\alpha,\beta\right)}\left(z\right)=\begin{cases}
\left(-\right)^{n+\lambda-5/2}G_{1,3}^{2,1}\left[2\kappa z\left|{\nicefrac{1}{2}\atop n,-n,\lambda-2}\right.\right], & \left(\lambda=\frac{5}{2},\frac{7}{2},\dots\right)\\
G_{2,4}^{2,2}\left[2\kappa z\left|{\lambda-\nicefrac{5}{2},\nicefrac{1}{2}\atop \lambda-\nicefrac{5}{2},n,-n,\lambda-2}\right.\right], & \left(\lambda\neq\frac{5}{2},\frac{7}{2},\dots\right).
\end{cases}\label{eq:MGZ18:I_n,k^a,b-2}
\end{equation}
We can now develop different representations for $\mathcal{I}_{n,\kappa}^{\left(\alpha,\beta\right)}\left(z\right)$
depending on $\lambda$.

\paragraph*{Case $\lambda=\nicefrac{5}{2},\nicefrac{7}{2},\dots$\@. }

First of all, writing $\lambda=\nicefrac{5}{2}+m$ $\left(m\geqslant0\right)$,
when $\lambda=\nicefrac{5}{2}$ we can employ (I.B11b)  and write
$I_{n,\kappa}^{\left(\alpha,\beta\right)}\left(z\right)$ as 
\[
\left.I_{n,\kappa}^{\left(\alpha,\beta\right)}\left(z\right)\right|_{\lambda=\frac{5}{2}}=\left(-\right)^{n}G_{1,3}^{2,1}\left[2\kappa z\left|{\nicefrac{1}{2}\atop n,-n,\nicefrac{1}{2}}\right.\right]=\left(-\right)^{n}G_{0,2}^{2,0}\left[2\kappa z\left|{-\atop n,-n}\right.\right]=\left(-\right)^{n}2K_{2n}\left(2\sqrt{2\kappa z}\right),
\]
where we have employed the representation of the modified Bessel function
$K_{\nu}\left(z\right)$ in terms of the $G$-function.\citep{Prudnikov90v3}

Then, using the differentiation formula 
\begin{equation}
\frac{d^{k}}{dz^{k}}\left\{ z^{-b_{q}}G_{p,q}^{m,n}\left[z\left|{\left(a_{p}\right)\atop \left(b_{q}\right)}\right.\right]\right\} =z^{-b_{q}-k}G_{p,q}^{m,n}\left[z\left|{\left(a_{p}\right)\atop \left(b_{q-1}\right),b_{q}+k}\right.\right]\quad\left(m<q\right),\label{eq:MGZ18:D^k-b_q+k}
\end{equation}
it is easy to conclude that 
\[
I_{n,\kappa}^{\left(\alpha,\beta\right)}\left(z\right)=2\left(-\right)^{m+n}\left.u^{\nicefrac{1}{2}+m}\frac{d^{m}}{du^{m}}\left[u^{-\nicefrac{1}{2}}K_{2n}\left(2\sqrt{u}\right)\right]\right|_{u=2\kappa z}.
\]

Finally, with the help of the identities\citep{Brychkov08}
\begin{align*}
K_{\nu+2n}\left(z\right) & =\sum_{k=0}^{n}\binom{n}{k}\left(\nu+n\right)_{k}\left(\frac{2}{z}\right)^{k}K_{\nu+k}\left(z\right), & \frac{\partial^{n}}{\partial z^{n}}\left[z^{\pm\nu/2}K_{\nu}\left(a\sqrt{z}\right)\right] & =\left(-\frac{a}{2}\right)^{n}z^{\left(\pm\nu-n\right)/2}K_{\nu\mp n}\left(a\sqrt{z}\right),
\end{align*}
we obtain 
\begin{equation}
\mathcal{I}_{n,\kappa}^{\left(\alpha,\beta\right)}\left(z\right)=\frac{\left(-\right)^{n}2\kappa}{\sqrt{\pi}\Gamma\left(\lambda-1\right)}\sum_{k=0}^{n}\sum_{\ell=0}^{\lambda-\nicefrac{5}{2}}\binom{n}{k}\binom{\lambda-\nicefrac{5}{2}}{\ell}\left(n\right)_{k}\left(\frac{1}{2}\right)_{\lambda-\frac{5}{2}-\ell}\frac{K_{k+\ell}\left(2\sqrt{2\kappa z}\right)}{\left(2\kappa z\right)^{\left(k-\ell\right)/2}},\;\left(\lambda=\frac{5}{2},\frac{7}{2},\dots\right).\label{eq:MGZ18:I_n,k^a,b-3}
\end{equation}

\paragraph*{Case $\lambda\protect\neq\nicefrac{5}{2},\nicefrac{7}{2},\dots$\@. }

In this case, the function $I_{n,\kappa}^{\left(\alpha,\beta\right)}\left(z\right)$
can be expressed as a combination of hypergeometric functions via
(I.B14), resulting
\begin{equation}
\begin{aligned}\mathcal{I}_{n,\kappa}^{\left(\alpha,\beta\right)}\left(z\right) & =-\frac{\left(-\right)^{n}\sqrt{\pi}\kappa}{\Gamma\left(\lambda-1\right)\cos\left(\pi\lambda\right)}\left[\Gamma\left(n+\frac{1}{2}\right)\left(2\kappa z\right)^{n}\pFRq 12\left({n+\nicefrac{1}{2}\atop 2n+1,3+n-\lambda};2\kappa z\right)\right.\\
 & \left.-\Gamma\left(\lambda-2\right)\left(2\kappa z\right)^{\lambda-\nicefrac{5}{2}}\pFRq 23\left({1,\lambda-2\atop \lambda-\nicefrac{3}{2}-n,\lambda-\nicefrac{3}{2}+n,\nicefrac{1}{2}};2\kappa z\right)\right],\:\left(\lambda\neq\frac{5}{2},\frac{7}{2},\dots\right),
\end{aligned}
\label{eq:MGZ18:I_n,k^a,b-4}
\end{equation}
where $\pFRq pq\left(\boldsymbol{\mathrm{a}};\boldsymbol{\mathrm{b}};z\right)$
is the regularized form of the $\pFq pq\left(\boldsymbol{\mathrm{a}};\boldsymbol{\mathrm{b}};z\right)$
hypergeometric function (see eq. 16.2.5 of Ref. \onlinecite{AskeyDaalhuis-Full-NIST10}).

\paragraph*{Derivative of $\mathcal{I}_{n,\kappa}^{\left(\alpha,\beta\right)}\left(z\right)$\@.}

When $\lambda$ is half-integer, we can employ the formula (\ref{eq:MGZ18:D^k-b_q+k})
again to obtain the recurrence relation
\begin{equation}
z\mathcal{I}_{n,\kappa}^{\left(\alpha,\beta\right)\prime}\left(z\right)=\left(\lambda-2\right)\mathcal{I}_{n,\kappa}^{\left(\alpha,\beta\right)}\left(z\right)-\left(\lambda-1\right)\mathcal{I}_{n,\kappa}^{\left(\alpha,\beta+1\right)}\left(z\right),\;\left(\lambda=\frac{5}{2},\frac{7}{2},\dots\right).\label{eq:MGZ18:I_n,k^a,b-lsi-deriv}
\end{equation}

On the other hand, when $\lambda\neq\nicefrac{5}{2},\nicefrac{7}{2},\dots$,
one can simply employ the formula\footnote{See \url{http://functions.wolfram.com/07.32.20.0005.01}.}
\[
\frac{d\hphantom{z}}{dz}\pFRq pq\left({\boldsymbol{\mathrm{a}}\atop \boldsymbol{\mathrm{b}}};z\right)=\left(\prod_{j=1}^{p}a_{j}\right)\pFRq pq\left({\boldsymbol{\mathrm{a}}+1\atop \boldsymbol{\mathrm{b}}+1};z\right)
\]
on (\ref{eq:MGZ18:I_n,k^a,b-4})\@. The derivation is straightforward
and will not be shown here.

\end{subequations}

\paragraph{Function $\mathcal{J}_{n,\kappa}^{\left(\alpha,\beta\right)}\left(z\right)$\@.}

Again, the $G$-function in the definition of $\mathcal{J}_{n,\kappa}^{\left(\alpha,\beta\right)}\left(z\right)$
given by (\ref{eq:MGZ18:X-asymp-xi}) is not defined when $\lambda-\nicefrac{5}{2}-n=1,2,\dots$\@.
Performing the same manipulations mentioned with regards to $\mathcal{I}_{n,\kappa}^{\left(\alpha,\beta\right)}\left(z\right)$,
we can write\begin{subequations}\label{eq:MGZ18:J_n,k^a,b-reps}
\begin{equation}
\mathcal{J}_{n,k,\kappa}^{\left(\alpha,\beta\right)}\left(x\right)=\frac{\kappa}{\sqrt{\pi}}J_{n,k,\kappa}^{\left(\alpha,\beta\right)}\left(x\right),\text{ where }J_{n,k,\kappa}^{\left(\alpha,\beta\right)}\left(x\right)=\begin{cases}
G_{1,3}^{2,1}\left[2\kappa x\left|{\nicefrac{1}{2}\atop \lambda-\nicefrac{5}{2}-k,n,-n}\right.\right], & \left(\lambda=\frac{5}{2},\frac{7}{2},\dots\right)\\
\left(-\right)^{k}G_{2,4}^{2,2}\left[2\kappa x\left|{\lambda-\nicefrac{5}{2},\nicefrac{1}{2}\atop \lambda-\nicefrac{5}{2},n,-n,\lambda-\nicefrac{5}{2}-k}\right.\right], & \left(\lambda\neq\frac{5}{2},\frac{7}{2},\dots\right).
\end{cases}\label{eq:MGZ18:J_n,k,k^a,b-1}
\end{equation}

\paragraph*{Case $\lambda=\nicefrac{5}{2},\nicefrac{7}{2},\dots$\@. }

Some of the formulae already obtained can be employed for the representation
of $\mathcal{J}_{n,k,\kappa}^{\left(\alpha,\beta\right)}\left(x\right)$\@.
For instance, we can write $\mathcal{J}_{n,k,\kappa}^{\left(\alpha,\beta\right)}\left(z\right)=\Gamma\left(\lambda-\nicefrac{3}{2}-k\right)\mathcal{H}_{n,\kappa}^{\left(\alpha,\beta-\nicefrac{1}{2}-k\right)}\left(z\right)$
and then use (I.24)\@. However, this is only valid for $\lambda-\nicefrac{5}{2}-k\geqslant0$\@. 

A better alternative is to identify the definition of $\mathcal{J}_{n,k,\kappa}^{\left(\alpha,\beta\right)}\left(x\right)$
with the function $F_{n,\theta}\left(z\right)$ in (\ref{eq:MGZ18:H_n,k^a,b-Integer})
and then evaluate the Mellin-Barnes integral using the residue theorem.
This procedure will result in a formula similar to (\ref{eq:MGZ18:Fng}),
with the caveat that now some of the poles of $\Gamma\left(\theta+s\right)$
are regularized by some of the poles of $\Gamma\left(n+1-s\right)$,
since $\Gamma\left(-r+\epsilon\right)/\Gamma\left(-p+\epsilon\right)\xrightarrow{\epsilon\to0}\left(-\right)^{r+p}p!/r!$
$\left(p,r=0,1,2,\dots\right)$\@. As a result, the first line in
(\ref{eq:MGZ18:Fng}) is replaced by 
\[
\epsilon_{\sigma}\frac{\left(\sigma-1\right)!\left(\nicefrac{1}{2}\right)_{\nu+\ell_{0}}z^{\nu+\ell_{0}}}{\left(\eta+\ell_{0}\right)!\ell_{0}!\left(1-\sigma\right)_{\ell_{0}}}\sum_{\ell=0}^{\sigma-1-\ell_{0}}\frac{\left(\nu+\nicefrac{1}{2}+\ell_{0}\right)_{\ell}z^{\ell}}{\left(1+\eta+\ell_{0}\right)_{\ell}\left(1-\sigma+\ell_{0}\right)_{\ell}\left(1+\ell_{0}\right)_{\ell}},
\]
where $\ell_{0}=\max\left(0,-\mu-\nu\right)$.

\paragraph*{Case $\lambda\protect\neq\nicefrac{5}{2},\nicefrac{7}{2},\dots$\@. }

Proceeding as per the same case for $\mathcal{I}_{n,\kappa}^{\left(\alpha,\beta\right)}\left(z\right)$,
we obtain
\begin{equation}
\begin{aligned}\mathcal{J}_{n,k,\kappa}^{\left(\alpha,\beta\right)}\left(z\right) & =\left(-\right)^{n+k+1}\frac{\sqrt{\pi}\kappa}{\cos\pi\lambda}\left[\Gamma\left(n+\frac{1}{2}\right)\left(2\kappa z\right)^{n}\pFRq 12\left({n+\nicefrac{1}{2}\atop 2n+1,n+\nicefrac{7}{2}-\lambda+k};2\kappa z\right)\right.\\
 & \left.-\Gamma\left(\lambda-2\right)\left(2\kappa z\right)^{\lambda-\nicefrac{5}{2}}\pFRq 23\left({1,\lambda-2\atop \lambda-\nicefrac{3}{2}-n,\lambda-\nicefrac{3}{2}+n,k+1};2\kappa z\right)\right],\quad\left(\lambda\neq\frac{5}{2},\frac{7}{2},\dots\right).
\end{aligned}
\label{eq:MGZ18:J_n,k,k^a,b-2}
\end{equation}

\paragraph*{Derivative of $\mathcal{J}_{n,k,\kappa}^{\left(\alpha,\beta\right)}\left(z\right)$\@.}

For half-integer $\lambda$, we employ de differentiation formula
\[
\frac{d}{dz}\left\{ z^{\sigma}G_{p,q}^{m,n}\left[z\left|{\left(a_{p}\right)\atop \left(b_{q}\right)}\right.\right]\right\} =\left(\sigma+b_{1}\right)z^{\sigma-1}G_{p,q}^{m,n}\left[z\left|{\left(a_{p}\right)\atop \left(b_{q}\right)}\right.\right]-z^{\sigma-1}G_{p,q}^{m,n}\left[z\left|{\left(a_{p}\right)\atop b_{1}+1,\left(b_{q-1}\right)}\right.\right]
\]
in order to obtain the recurrence relation
\begin{equation}
x\mathcal{J}_{n,k,\kappa}^{\left(\alpha,\beta\right)\prime}\left(x\right)=\left(\lambda-\frac{5}{2}-k\right)\mathcal{J}_{n,k,\kappa}^{\left(\alpha,\beta\right)}\left(x\right)-\mathcal{J}_{n,k,\kappa}^{\left(\alpha,\beta+1\right)}\left(x\right),\;\left(\lambda=\frac{5}{2},\frac{7}{2},\dots\right).\label{eq:MGZ18:J_n,k,k^a,b-lsi-deriv}
\end{equation}

When $\lambda\neq\nicefrac{5}{2},\nicefrac{7}{2},\dots$, we derive
directly (\ref{eq:MGZ18:J_n,k,k^a,b-2}).

\end{subequations}

\end{widetext}

\section{Numerical solutions of the dispersion equation\label{sec:MGZ18:Numerical_sols}}

In this section we will present some numerical solutions of the dispersion
equation (\ref{eq:MGZ18:DE-BK-OAW})\@. For the implementation of
the computer code, we employed several of the properties presented
in Papers I and II, as well as in section \ref{sec:MGZ18:Kappa_plasma_functions}.

The bulk of the code was written in Modern Fortran,\citep{Metcalf+11}
but several key components were made possible thanks to the multiple-precision
libraries \texttt{MPMath}\citep{mpmath} and \texttt{Arb},\citep{Johansson17/08}
respectively written in Python and C\@. The C functions are accessed
from Fortran with the Application Programmer's Interface (API) present
in the standard of the language, whereas the Python modules are acessed
via the P/C API.\citep{Python-C-API}

In this work we will present only some representative solutions of
equation (\ref{eq:MGZ18:DE-BK-OAW})\@. A more detailed and comprehensive
analysis of the oblique firehose instability occurring in kappa plasmas
will be presented in a future publication. In order to reduce the
number of symbols employed in the discussion, we adopt the following
normalized forms,
\begin{align*}
q_{\parallel(\perp)} & =\frac{k_{\parallel(\perp)}v_{A}}{\Omega_{i}}, & z_{r(i)} & =\frac{\omega_{r(i)}}{\Omega_{i}},
\end{align*}
where $\omega=\omega_{r}+i\omega_{i}$.

\begin{figure}
\noindent \begin{centering}
\includegraphics[width=1\columnwidth]{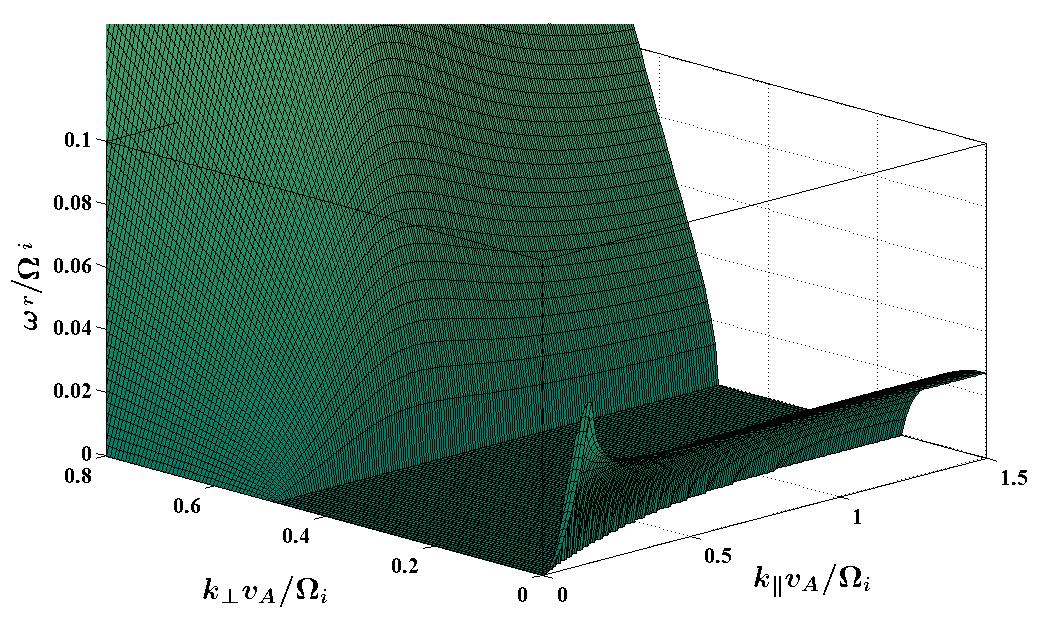}
\par\end{centering}
\noindent \begin{centering}
\includegraphics[width=1\columnwidth]{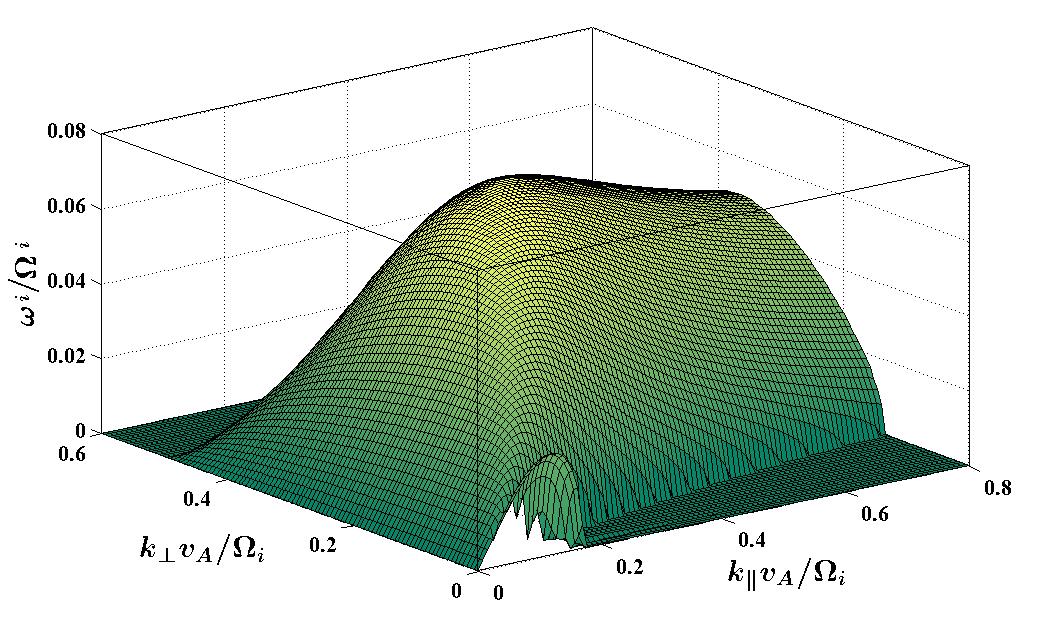}
\par\end{centering}
\caption{Plots of the real frequency (top panel) and of the growth rate (bottom
panel) and of the unstable mode versus wave number for $\beta_{e}=2$
$\left(T_{\parallel e}=T_{\perp e}\right)$, $\beta_{\parallel i}=3$,
$\beta_{\perp i}=0.8$ and $\kappa_{e}=\kappa_{i}=5$.\label{fig:MGZ18:gr-3d}}
\end{figure}

Figure \ref{fig:MGZ18:gr-3d} shows a typical solution of the dispersion
equation (\ref{eq:MGZ18:DE-BK-OAW})\@. We plotted only the normalized
values of the dispersion relation (top panel) and of the growth rate
(bottom panel) of the unstable mode $\left(\omega_{i}/\Omega_{i}\right)$
versus the normalized parallel $\left(k_{\parallel}v_{A}/\Omega_{i}\right)$
and  perpendicular $\left(k_{\perp}v_{A}/\Omega_{i}\right)$ components
of the wave vector. The physical parameters used in figure \ref{fig:MGZ18:gr-3d}
are the following: the electron VDF is isotropic, with plasma beta
$\beta_{e}=\beta_{\parallel e}=\beta_{\perp e}=2$; the ion VDF is
anisotropic, with $\beta_{\parallel i}=3$ and $\beta_{\perp i}=0.8$,
which corresponds to a temperature ratio of $T_{\parallel i}/T_{\perp i}=3.75$
or to an anisotropy parameter $A_{i}=1-T_{\perp i}/T_{\parallel i}=0.733$\@.
These are typical parameters to excite the firehose instability. Additionaly,
both VDFs are superthermal, with $\kappa_{e}=\kappa_{i}=5$\@. 

Figure \ref{fig:MGZ18:gr-3d} is similar to Fig. 4 of Yoon \emph{et
al}\@.\citep{YoonWuAssis93/07} In the bottom panel, one can observe
that the growth rate of the instability is split in two branches:
on the plane $k_{\perp}=0$ the instability is restricted to the range
$0\leqslant k_{\parallel}v_{A}/\Omega_{i}\lesssim0.16$\@. This branch
rapidly vanishes as $k_{\perp}$ grows, while the other branch of
the instability displays a growing behavior which climbs to a maximum
$z_{i}\approx0.07$ at $\left(q_{\parallel},q_{\perp}\right)\approx\left(0.37,0.32\right)$
and then gradually vanishes as well as $q_{\perp}$ grows, but falls
slowly along the $q_{\parallel}$ direction. This branch of the instability
was called the \emph{oblique firehose} by Yoon \emph{et al.}\citep{YoonWuAssis93/07}
and the \emph{Alfvén firehose} by Hellinger and Matsumoto\@.\citep{HellingerMatsumoto00/05}
Another noticeable aspect is that the maximum growth rate of the oblique
firehose is substantially larger than the maximum growth-rate of the
parallel branch $\left(z_{i}\approx0.02\right)$.

Over the spectral range where the oblique firehose is operative, one
observes, in the top panel of figure \ref{fig:MGZ18:gr-3d}, that
the real part is zero; \emph{i.e.}, the oblique firehose instability
occurs in a nonpropagating mode. This characteristic was pointed out
by Refs. \onlinecite{YoonWuAssis93/07,HellingerMatsumoto00/05} and
is also valid for a kappa plasma.

\begin{figure}[b]
\begin{centering}
\includegraphics[width=1\columnwidth]{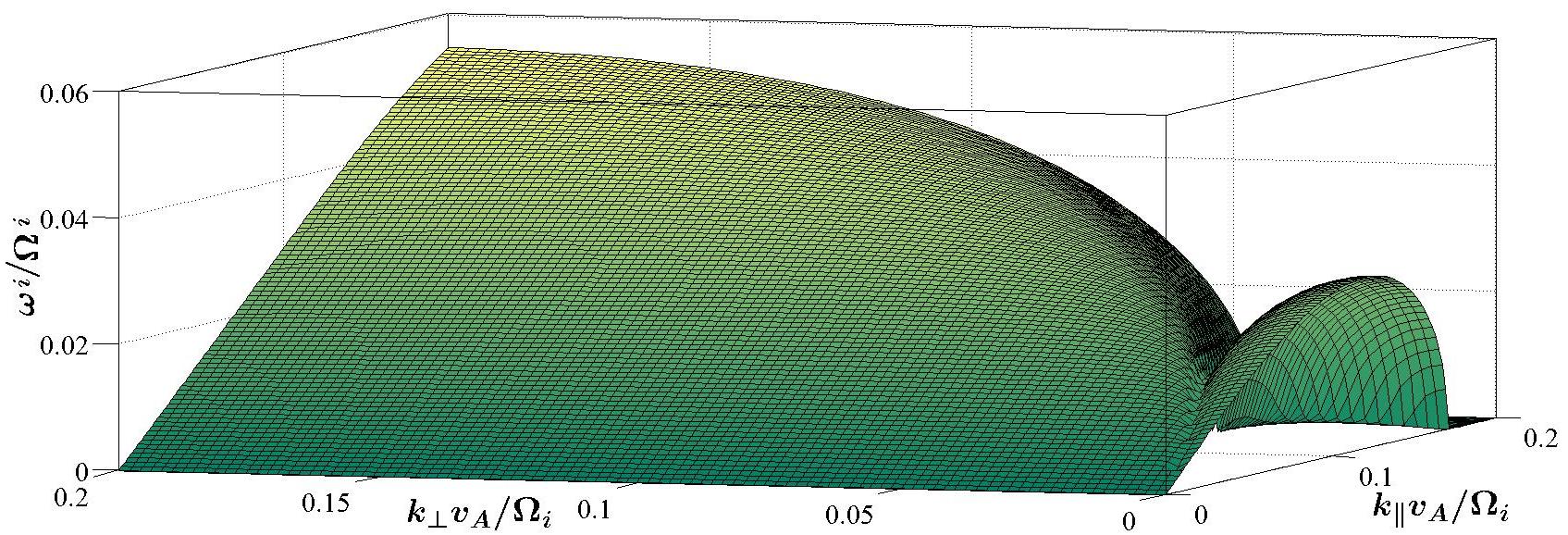}
\par\end{centering}
\caption{Transition between the parallel firehose instability and the oblique
firehose. All other parameters as in Fig.  \ref{fig:MGZ18:gr-3d}\@.\label{fig:MGZ28:gr-parallel-oblique}}
\end{figure}

The transition between the parallel and oblique branches of the instability
can be seen in greater detail in figure \ref{fig:MGZ28:gr-parallel-oblique},
which is an inset of the bottom panel of figure \ref{fig:MGZ18:gr-3d}\@.
One can clearly observe the shooth transition between either branch
of the instability, with the parallel branch confined in the range
$0\leqslant q_{\perp}\lesssim0.03$, $0\leqslant q_{\parallel}\lesssim0.16$
and the oblique brach operative for $q_{\perp}\gtrsim0.03$\@.

The surface plots in figures \ref{fig:MGZ18:gr-3d} and \ref{fig:MGZ28:gr-parallel-oblique}
show the characteristics of the firehose instability for a single
combination of electron/ion kappa parameters. If one wishes to analyze
the dependency of the instability with different values for the kappas,
3D surface plots are not adequate. Instead, we will take the coordinates
of the maximum growth rate for a Maxwellian plasma, which are $\left(q_{\parallel},q_{\perp}\right)\approx\left(0.41,0.37\right)$
for the parameters in figure \ref{fig:MGZ18:gr-3d}, fix either $k_{\parallel}$
or $k_{\perp}$ and then plot the growth rates along the other component
of the wave vector for several different values of $\kappa_{e}$,
$\kappa_{i}$\@.

\begin{figure}
\begin{centering}
\includegraphics[width=1\columnwidth]{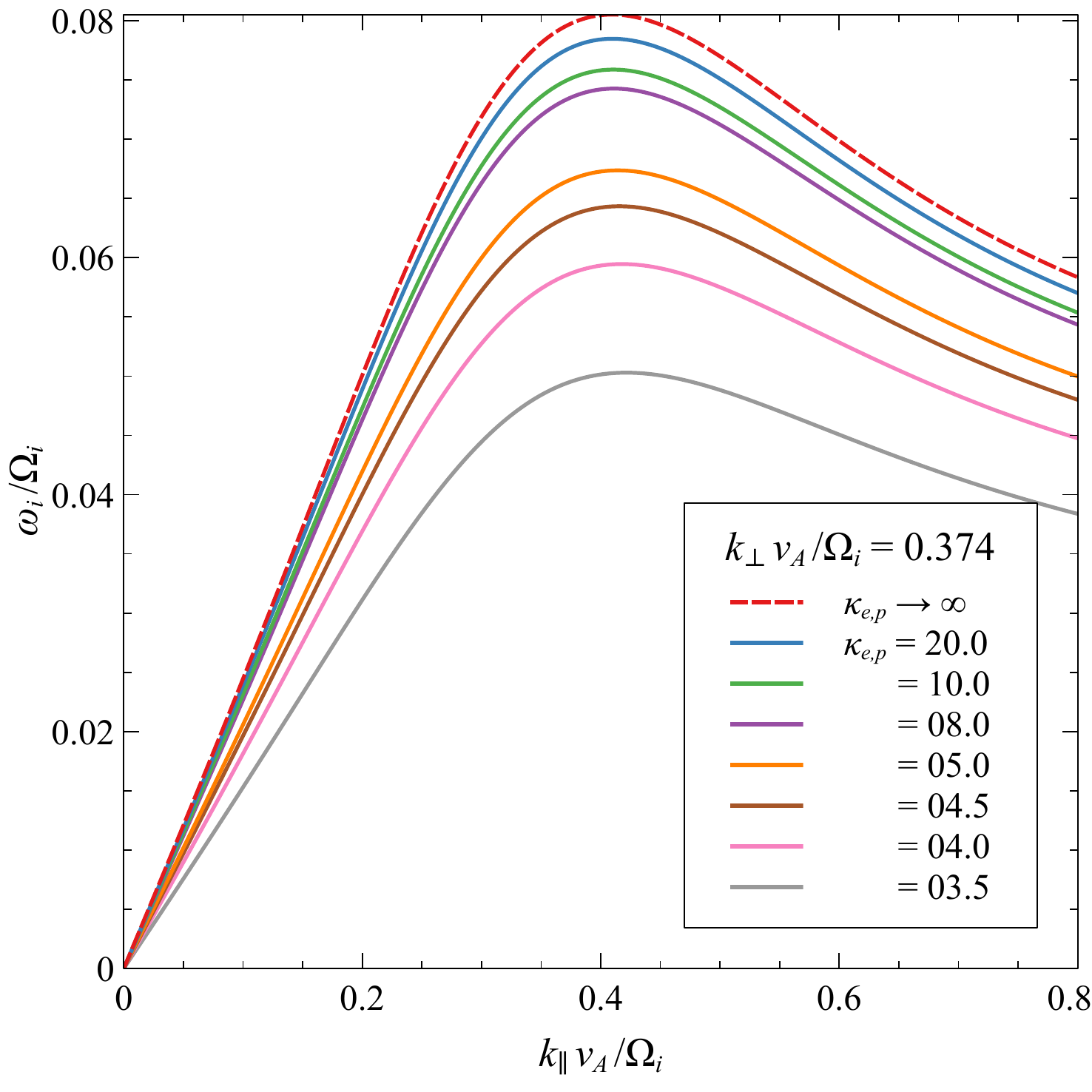}
\par\end{centering}
\begin{centering}
\includegraphics[width=1\columnwidth]{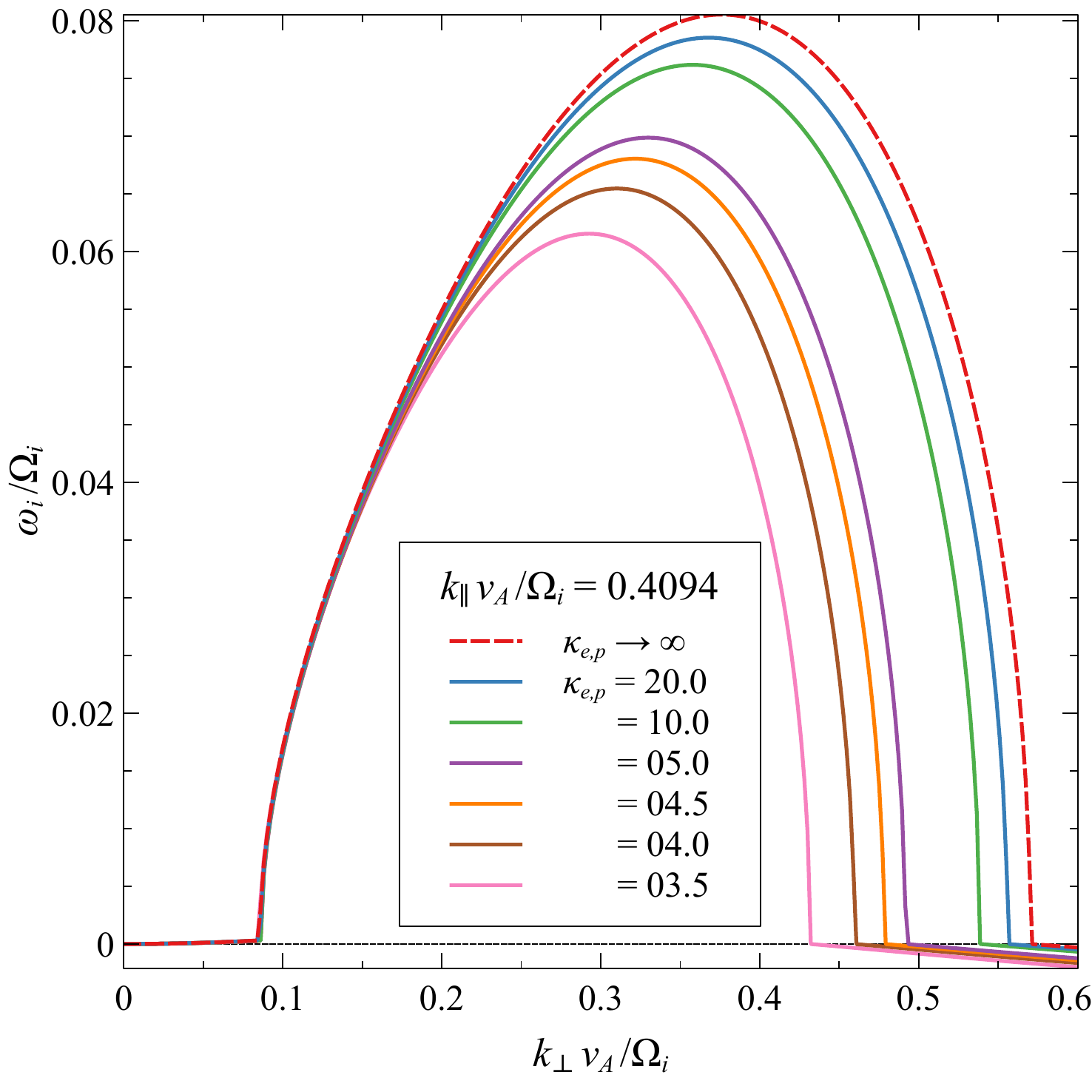}
\par\end{centering}
\caption{Plots of $\omega_{i}/\Omega_{i}$ for several values of $\kappa_{e}=\kappa_{i}$\@.
Top panel: varying $k_{\parallel}$, keeping $q_{\perp}=0.374$ fixed.
Bottom panel: varying $k_{\perp},$ keeping $q_{\parallel}=0.4094$
fixed. All other parameters are the same as in figure \ref{fig:MGZ18:gr-3d}.\label{fig:MGZ18:gr-1D}}
\end{figure}

Proceeding in this way, we obtain the results shown in figure \ref{fig:MGZ18:gr-1D}\@.
In the present analysis, we will consider the particular choice of
$\kappa_{e}=\kappa_{i}$\@. A more realistic and comprehensive analysis
will be presented in a future publication. In the top panel of figure
\ref{fig:MGZ18:gr-1D}, we show the dependence of $\omega_{i}$ with
$q_{\parallel}$ for a fixed $q_{\perp}=0.374$\@. The dashed curve
is the solution of the Maxwellian limit of the dispersion equation
(\ref{eq:MGZ18:DE-BK-OAW}), obtained directly from the expressions
for a bi-Maxwellian VDF\@. The blue curve of Fig. \ref{fig:MGZ18:gr-1D}(top),
on the other hand, corresponds to the solution of (\ref{eq:MGZ18:DE-BK-OAW})
with $\kappa_{e}=\kappa_{i}=20$\@. As expected, this case is already
close to the pure Maxwellian plasma environment. As the kappa indices
decrease, the maximum growth rate along $q_{\parallel}$ also drops
(from $z_{i,\mathrm{max}}\approx0.08$ for $\kappa\to\infty$ to $z_{i,\mathrm{max}}\approx0.062$
for $\kappa=3.5$), with the value of $q_{\parallel,\mathrm{max}}=0.41$
approximately the same for all kappas.

A different behavior is observed along $q_{\perp}$\@. The bottom
panel of Fig. \ref{fig:MGZ18:gr-1D} again shows that the case $\kappa_{e}=\kappa_{i}=20$
is already close to the Maxwellian limit and that $z_{i,\mathrm{max}}$
drops as the kappas are reduced, with the same variation observed
in the top panel. However, along the perpendicular direction one can
observe some distinguishing features not aparent in the top panel.
First of all, in the small $k_{\perp}$ region $\left(q_{\perp}\lesssim0.17\right)$,
corresponding to the ``small gyroradius'' case, the growth rate
remains roughly independent of $\kappa$, with the limiting situation
that at $q_{\perp}=0$ the solution is exactly the same as in the
Maxwellian case. On the other hand, for $q_{\perp}\gtrsim0.17$ the
growth rate becomes dependent on $\kappa$, in such a way that not
only the value of $z_{i,\mathrm{max}}$ reduces with $\kappa$, but
the spectral range of the instability in the perpendicular direction
reduces as well. Hence, these results suggest that for moderate values
of the gyroradius, the oblique firehose instability is strongly dependant
on the kappa parameter.

\begin{figure}
\begin{centering}
\includegraphics[width=1\columnwidth]{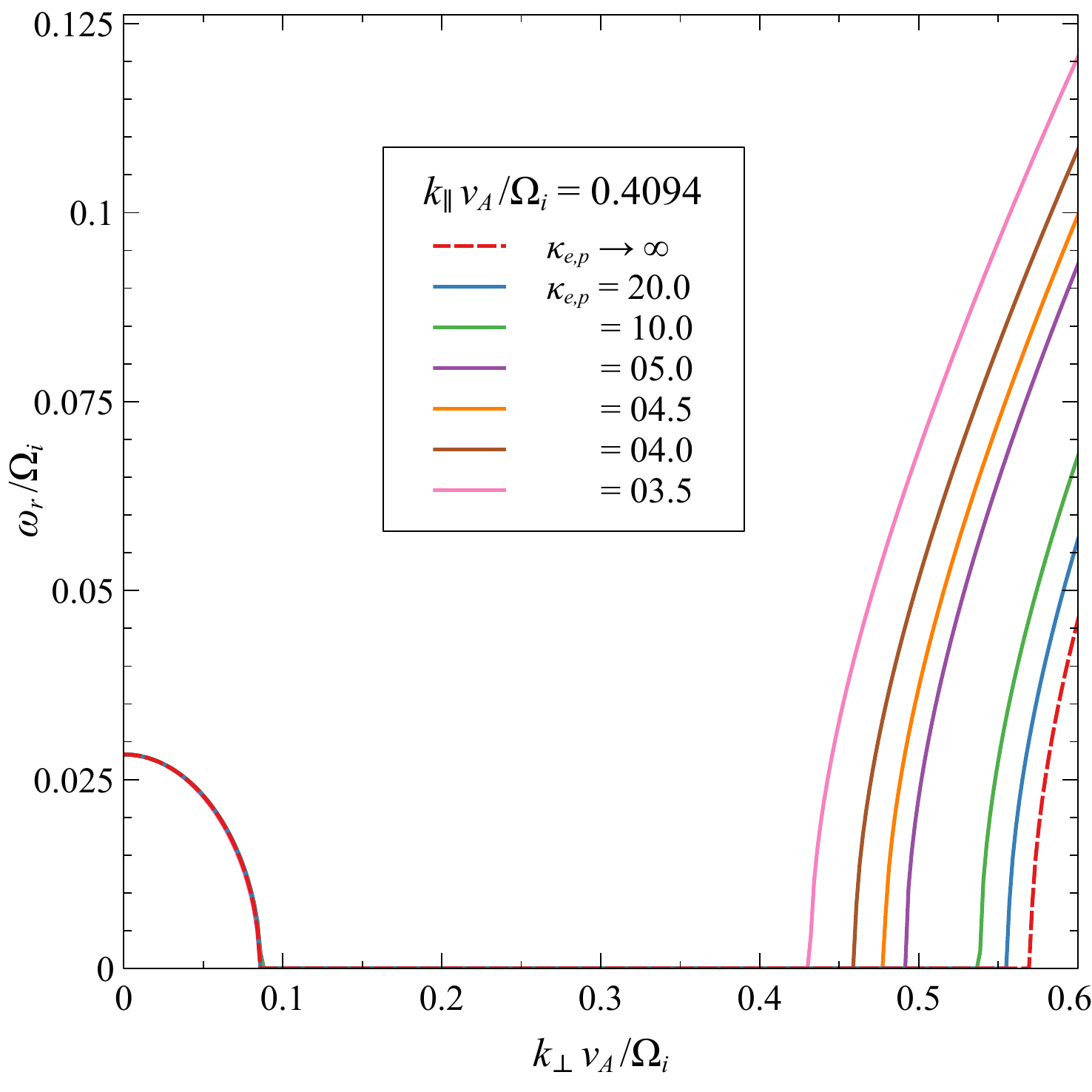}
\par\end{centering}
\caption{Plots of $\omega_{r}/\Omega_{i}$ for several values of $\kappa_{e}=\kappa_{i}$,
varying $k_{\perp}$ with $q_{\parallel}=0.4094$ fixed. All other
parameters are the same as in figure \ref{fig:MGZ18:gr-3d}.\label{fig:MGZ18:gr-1D-Re}}
\end{figure}

The same behavior is displayed by the real part of the unstable mode
as a function of $k_{\perp}$, as can be seen in Figure \ref{fig:MGZ18:gr-1D-Re}\@.
In the low gyroradious limit, the wave refracts as in a Maxwellian
plasma, then the mode becomes nonpropagating throughout the unstable
spectral range and finally becomes propagating again right at the
point where the instability disapears and is replaced by damping.
The value of $k_{\perp}$ where the mode ceases to be unstable is
dependent on the kappa parameter, with the nonpropagating, unstable
spectral range consistently reducing with $\kappa_{e}=\kappa_{i}$.

\begin{figure}
\begin{centering}
\includegraphics[width=1\columnwidth]{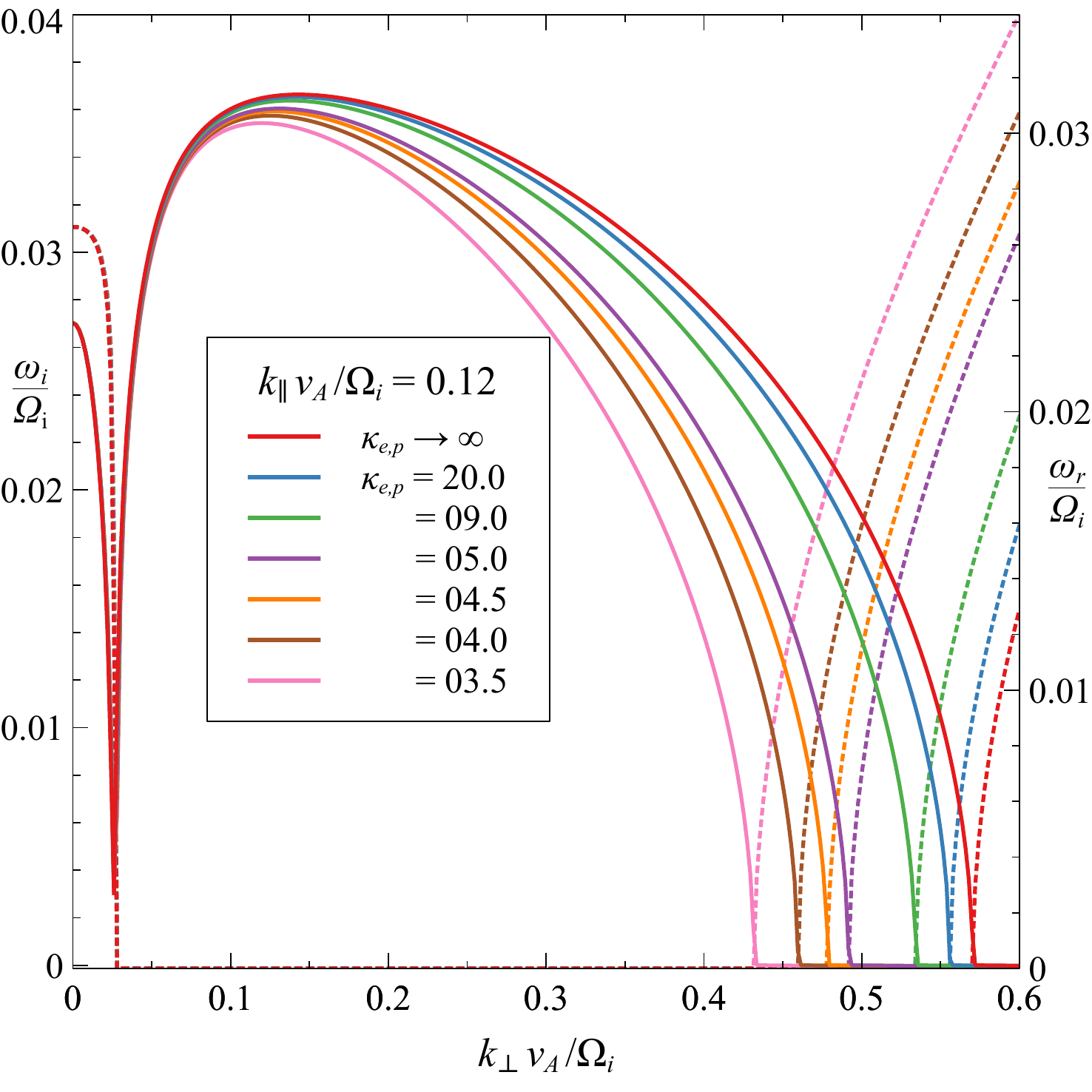}
\par\end{centering}
\caption{Plots of $\omega_{i}/\Omega_{i}$ (continuous lines) and $\omega_{r}/\Omega_{i}$
(dotted lines) as functions of $q_{\perp}$ for fixed $q_{\parallel}=0.12$
and various values of $\kappa_{e}=\kappa_{i}$\@. All other parameters
as in Fig. \ref{fig:MGZ18:gr-3d}.\label{fig:MGZ18:gr-1D-2}}
\end{figure}

As a final result, figure \ref{fig:MGZ18:gr-1D-2} shows both the
normalized growth rate (continuous lines) and the normalized real
part (dotted lines) of the unstable mode for a fixed value of $q_{\parallel}$
and varying $q_{\perp}$\@. Now, the parallel component of the wavenumber
is fixed to $k_{\parallel}v_{A}/\Omega_{i}=0.12$, which corresponds
to the maximum growth rate of the (quasi) parallel branch of the firehose
instability. The growth rates displayed by the figure clearly show
the transition from the quasi-parallel to the oblique branches of
the instability, which occurs at $q_{\perp}\approx0.026$ for all
values of the kappa parameter. We observe the same behavior displayed
by Figs. \ref{fig:MGZ18:gr-1D} and \ref{fig:MGZ18:gr-1D-Re}: the
quasi-parallel branch is almost independent on $\kappa_{e}$ and $\kappa_{i}$,
whereas the oblique mode shows an evident dependence on the kappas.
We again observe that not only the maximum growth rate is reduced
with kappa, but so does also the unstable spectral range. 

The real part of the unstable mode also repeats the same pattern observed
in the previous figures: the quasi-parallel branch of the instability
is convective, with nonzero phase velocity that is almost independent
on the kappa values. On the other hand, the oblique branch is nonpropagating
throughout the unstable spectral range and acquires a nonzero phase
velocity when the instability disappears, being replaced by a very
small damping coefficient.

As a final remark, we mention again that a more complete treatment
will show that the quasi-parallel branch of the firehose instability
does indeed depend on $\kappa_{i}$\@. However, this does not invalidate
the present treatment, since the main objective was to study the effect
of the superthermal nature of the electron and ion distribution functions
on the oblique firehose instability, which does depend on the kappas.

\section{Conclusions\label{sec:MGZ18:Conclusions}}

We presented the derivation of a dispersion equation that describes
the oblique firehose instability excited in an electron-ion plasma
depending on the wave vector, the parallel and perpendicular electron
and ion beta parameters, and on the kappa parameters of the electron
and ion velocity distribution functions. 

In order to implement the numerical solution of the dispersion equation,
several new mathematical properties of the kappa plasma special functions
were obtained, which complement the formalism already derived in previous
publications.

Employing values of the physical parameters that are relevant to space
plasma conditions, some solutions of the dispersion equation were
shown. The results show that both the maximum growth rate of the instability
and its spectral range depend on the superthermal nature of the $\kappa$
distributions, with both properties roughly displaying a reduction
with the values of $\kappa_{e}=\kappa_{i}$\@. 

A more comprehensive and complete analysis of the oblique firehose
instability was not reported here, due to the length of the paper
demanded by the mathematical expressions. This task will be carried
out in future publications, not only for the firehose instability
but also for other relevant instabilities occurring in arbitrary angles,
polarization and frequency ranges.
\begin{acknowledgments}
The authors acknowledge support provided by Conselho Nacional de Desenvolvimento
Científico e Tecnológico (CNPq), grants No. 304363/2014-6 and 307626/2015-6\@.
This study was financed in part by the Coordenação de Aperfeiçoamento
de Pessoal de Nível Superior - Brasil (CAPES) - Finance Code 001.
\end{acknowledgments}

\bibliographystyle{aipnum4-1}
\bibliography{revabr,plasma,splasma,matematica,comp,physics}

\end{document}